\documentclass[preprint,aps,superscriptaddress]{revtex4}

\usepackage{graphicx}
\usepackage{graphics}
\usepackage{natbib}

\usepackage{amsmath}
\usepackage{amssymb}

\newcommand{\bb}{\boldsymbol{b}}
\newcommand{\bhat}{\boldsymbol{\hat{b}}}
\newcommand{\bu}{\boldsymbol{u}}
\newcommand{\bup}{\boldsymbol{u}_p}
\newcommand{\buk}{\boldsymbol{u_k}}

\newcommand{\pr}{\partial}
\newcommand{\bnabla}{\boldsymbol{\nabla}}
\newcommand{\bk}{\boldsymbol{k}}

\begin{document}

 \title{Reduction of compressibility and parallel transfer by Landau damping in turbulent magnetized plasmas}

\author{P. Hunana} 
\email{hunana@oca.eu}
\author{D. Laveder}
\author{T. Passot}
\author{P. L. Sulem}

\affiliation{Universit\'e de Nice Sophia Antipolis, CNRS, Observatoire de la C\^ote d'Azur, \\ 
BP 4229 06304, Nice Cedex 4, France\\}

\author{D. Borgogno}
\affiliation{Dipartimento di Energetica, Politecnico di Torino, corso Duca degli Abruzzi 24, 10138 Torino, Italy}

\date{\today}

\begin{abstract}
Three-dimensional numerical simulations of decaying turbulence in a magnetized plasma 
are performed using a so-called FLR-Landau fluid model
which incorporates linear Landau damping and finite Larmor radius (FLR) corrections. 
It is shown that compared to simulations of compressible Hall-MHD, linear Landau damping 
is responsible for significant damping of magnetosonic waves, which is consistent with the linear kinetic theory. 
Compressibility of the fluid and parallel energy cascade along the ambient magnetic field are also significantly inhibited 
when the beta parameter is not too small. 
In contrast with Hall-MHD, the FLR-Landau fluid model can therefore correctly describe
turbulence in collisionless plasmas such as the solar wind, providing an interpretation for its 
nearly incompressible behavior.  
\end{abstract}

\maketitle


\section{Introduction}
Hydrodynamics and Magnetohydrodynamics (MHD) are the central descriptions used to study turbulence in the solar wind 
and in a wide range of natural systems. Specifically for the solar wind, MHD description yielded a great success 
in our understanding of observational data (e.g. see reviews by Goldstein et al. \cite{GoldsteinRoberts1995}, 
Tu and Marsch \cite{TuMarsch1995}, Bruno and Carbone 
\cite{BrunoCarbone2005}, Horbury et al. \cite{Horbury2005}, Marsch \cite{Marsch2006}, Ofman \cite{Ofman2010}). 
Observational studies show that the solar wind is typically found to be 
only weakly compressible (e.g. see Matthaeus et al. \cite{MatthaeusKlein1991}, 
Bavassano and Bruno \cite{BavassanoBruno1995} and the reviews cited above) and usual
turbulence models which predict the energy spectra are derived 
in the framework of an incompressible MHD description 
(Iroshnikov \cite{Iroshnikov1963}, Kraichnan \cite{Kraichnan1965}, 
Goldreich and Sridhar \cite{GoldreichSridhar1995,GoldreichSridhar1997}, Galtier et al. 
\cite{GaltierNazarenko2000,Galtier2002}, Boldyrev \cite{Boldyrev2005,Boldyrev2006}, 
Lithwick et al. \cite{Lithwick2007}, Chandran \cite{Chandran2008}, Perez and Boldyrev \cite{PerezBoldyrev2009}, 
Podesta and Bhattacharjee \cite{PodestaBhattacharjee2010}, 
see also reviews by Cho et al. \cite{ChoLazVish2003}, Zhou et al. \cite{ZhouMatthaeusDmitruk2004}, 
Galtier \cite{Galtier2009} and Sridhar \cite{Sridhar2010}).
Theoretical models which describe the radial evolution of spatially averaged solar wind quantities
are also usually developed in the framework of incompressible MHD formulated in Els\"asser variables 
(e.g. Zhou and Matthaeus \cite{ZhouMatthaeus1989,ZhouMatthaeus1990a,ZhouMatthaeus1990b}, 
Marsch and Tu \cite{MarschTu1989}, Zank et al. \cite{ZMS1996}, Smith et al. \cite{Smith2001}, 
Matthaeus et al. \cite{Matthaeus2004}, Breech et al. \cite{Breech2008}). 
It is however well known that the solar wind is not completely incompressible and many 
phenomena which require compressibility are observed in the solar wind, such as 
the evolution of density fluctuations (e.g. Spangler and Armstrong \cite{Spangler1990}, 
Armstrong et al. \cite{ArmstrongColes1990}, Coles et al. \cite{Coles1991}, 
Grall et al. \cite{Grall1997}, Woo and Habbal \cite{WooHabbal1997}, Bellamy et al. \cite{Bellamy2005},
Wicks et al. \cite{Wicks2009}, Telloni et al. \cite{Telloni2009}), 
magnetic holes, solitons and mirror mode structures (e.g. Winterhalter et al. \cite{Winterhalter1994},
Fr\"anz et al. \cite{Franz2000}, Stasiewicz et al. \cite{Stasiewicz2003}, Stevens and Kasper \cite{StevensKasper2007})
or strong temperature anisotropies which trigger, and are limited by, micro-instabilities such 
as mirror and fire-hose   
(e.g. Gary et al. \cite{Gary2001}, Kasper et al. \cite{Kasper2002}, Hellinger et al. \cite{Hellinger2006},
Matteini et al. \cite{Matteini2007}, Bale et al. \cite{Bale2009}).
The importance of compressibility was stressed by Carbone et al. \cite{Carbone2009},
who compared the observational data with the energy flux scaling laws of 
Politano and Pouquet \cite{Politano_GRL1998}, which are exact relations of incompressible MHD. 
Carbone et al. determined that the scaling relations can better fit the data
if the relations are phenomenologically modified to account for compressibility.  
The incorporation of weakly compressional density fluctuations was partially addressed by 
so-called nearly incompressible models, which expand the compressible equations 
with respect to small sonic Mach number (Matthaeus and Brown \cite{MB1988}, Zank and Matthaeus \cite{ZM1993,ZM1991,ZM1990}) 
and which were recently formulated in the presence of a static large-scale inhomogeneous background
(Hunana and Zank \cite{Hunana2010}, Hunana et al. \cite{Hunana2006,HunanaJGR2008}, 
see also Bhattacharjee et al. \cite{Bhattacharjee1998}). These models however specifically assume, and cannot explain, 
why the solar wind is only weakly compressible. 
The theoretical compressible MHD models
developed in the wave turbulence formalism (e.g. Kuznetsov \cite{Kuznetsov2001}, Chandran \cite{Chandran2005})
also cannot address this issue.

Describing the solar wind with fully compressible MHD or Hall-MHD formalisms yields several problems. 
Most importantly, these compressible descriptions introduce sound waves and slow magnetosonic waves. As elaborated
by Howes \cite{Howes2009}, slow magnetosonic waves are strongly damped by Landau damping 
in the kinetic Maxwell-Vlasov description. The presence of fast and slow magnetosonic waves 
naturally implies higher level of compressibility and overestimates the parallel energy transfer. 
Also, numerical simulations of compressible Hall-MHD performed by Servidio et al. \cite{Servidio2007}
in the context of the magnetopause boundary layer showed that compared to the usual turbulence 
in compressible MHD simulations, which consists of Alfv\'en waves, 
the Hall term is responsible for decoupling of magnetic and velocity field fluctuations. In compressible 
Hall-MHD regime, Servidio et al. observed spontaneous generation of magnetosonic waves which
transform to a regime of quasi perpendicular ``magnetosonic turbulence''. This is 
in contrast with observational studies which typically show that turbulence in the solar wind 
predominantly consists of quasi perpendicular (kinetic) Alfv\'en waves 
(e.g., Bale et al. \cite{Bale2005}, Sahraoui et al. \cite{Sahraoui2009,Sahraoui2010}).  
Finally, all usual MHD or Hall-MHD models cannot address how the energy
is actually dissipated at small scales. It is well known that solar wind plasma is almost collisionless and 
therefore the classical von Karman picture of energy being dissipated via viscosity is not applicable 
to the solar wind. It is evident that new and more realistic models have to be introduced which can overcome some of these drawbacks.

The most realistic approach is of course 
the fully kinetic Vlasov-Maxwell description. It is however analytically intractable, 
and even the biggest kinetic simulations cannot resolve the large-scale turbulence dynamics. 
Two leading approaches appear to be promising to substitute MHD in describing solar 
wind turbulence : Gyrokinetics and Landau fluids.  
Gyrokinetics (e.g. Schekochihin et al. \cite{Schekochihin2009}, Howes et al. \cite{Howes2006}) 
was originally developed for simulations of fusion in tokamaks. It is a kinetic-like description, which 
averages out the gyro-rotation of particles around a mean magnetic field and therefore makes the 
kinetic description more tractable, mostly by eliminating fast time scales. Derived directly from the kinetic theory, gyrokinetics has a 
crucial advantage of being asymptotically correct. Landau fluid description, on the other hand, is a fluid-like extension 
of compressible Hall-MHD, in which wave dissipation is incorporated kinetically by the modeling of linear Landau damping, 
thus retaining a realistic sink of energy. Other linear kinetic effects such as finite Larmor radius corrections are 
also incorporated. 

The simplest Landau fluid closure was considered by Hammett and Perkins \cite{HammettPerkins1990} 
and the associated dispersion relations were numerically explored by Jayanti, Goldstein and Vi\~nas \cite{Jayanti1998}.
The Landau fluid model was further advanced by Snyder, Hammett and Dorland \cite{Snyder1997},
who considered the largest MHD scales, starting from the guiding center kinetic equation. 
The approach was reconsidered and refined to its present form with incorporation of Hall term 
and finite Larmor radius (FLR) corrections 
by Passot, Sulem, Goswami and Bugnon \cite{PassotSulem2007,PassotSulem2003b,Goswami2005,Bugnon2004}. 
This Landau fluid approach starts with the Vlasov-Maxwell equations and  
derives nonlinear evolution equations for density, velocity and gyrotropic pressures.
In the simplest formulation the model is closed at the level of heat fluxes 
by matching with the linear kinetic theory in the low frequency limit. Kinetic expressions usually contain  
the plasma dispersion function which is not suitable for fluid-like simulations. 
Landau fluid closure is therefore performed in a way as to minimize occurrences of this plasma dispersion function 
and, where not possible, this function is replaced by a Pad\'e approximant.  
This eliminates the time non-locality and also results in the presence of a Hilbert transform 
with respect to the longitudinal coordinate 
(in the direction of the ambient magnetic field) which in the fluid formalism is associated with linear Landau damping.  
Further details about the development of Landau fluid models are thoroughly discussed in the 
papers cited above. The Landau fluid approach should however be contrasted with the more classical gyrofluid models 
(e.g. Dorland and Hammett \cite{DorlandHammett1993}, Brizard \cite{Brizard1992}, Scott \cite{Scott2010}), 
which are derived by taking fluid moments of the gyrokinetic equation and for which a similar closure scheme 
is applied afterwards.

The Landau fluid approach has the following advantages. In contrast with Hall-MHD, 
it contains separate equations for parallel and perpendicular pressures 
and heat fluxes. It therefore allows for the development of temperature anisotropy, which is observed in the 
solar wind. Noticeably, in contrast with gyrokinetics or gyrofluid models, the Landau fluid model does not average 
out the fast waves. Compared with these approaches, it also has an advantage in
that the final equations including the FLR corrections are written
for the usual quantities measured in the laboratory frame. 
Existing spectral MHD and Hall-MHD codes can therefore be modified 
to Landau fluid description relatively easily.  
Also importantly, even though gyrokinetics is a reduced kinetic description, it is still 5-dimensional and therefore 
naturally quite difficult to compute. While current largest numerical simulations of gyrokinetics 
(Howes et al. \cite{HowesPRL2008}) require thousands CPU-cores for a fluid-like $128\times 64^2$ resolution, 
the FLR-Landau fluid model requires computational power only slightly larger than the usual Hall-MHD simulations. The 
results presented here, which employ a resolution of $N=128$ grid points in all three directions, 
were calculated using 32 CPU-cores. 
 
Landau fluid models can be developed with several levels of complexity. 
For these first Landau fluid simulations of three dimensional turbulence,
we use a simplified version of the most general Landau fluid model \cite{PassotSulem2007}, where we constrain
ourselves to isothermal electrons and leading order corrections in terms of the ratio of the ion Larmor radius to the 
considered scales. 
A similar model was used by Borgogno et al. \cite{Borgogno2009}, who studied the dynamics of 
parallel propagating Alfv\'en waves in a medium with an inhomogeneous density profile. 
They numerically showed that the observed Alfv\'en wave filamentation and 
later transition to the regime of dispersive phase mixing is consistent with particle-in-cell simulations. 
In this paper we concentrate on freely decaying turbulence and compare these to simulations of compressible Hall-MHD. 
Numerical integration of the full Landau fluid model in one space dimension was presented by Borgogno et al. \cite{Borgogno2007},
who investigated the dynamics very close to the mirror instability threshold and showed the presence of magnetic holes.
Results considering quasi-transverse one-dimensional propagation in the full Landau fluid model are
presented in \cite{Laveder1D,Laveder_prep}.


\section{The model and its numerical implementation}

Considering a neutral bi-fluid consisting of protons (ions) and isothermal electrons,
the Landau fluid model consists of evolution equations for proton density $\rho=m_p n$ 
(where $m_p$ is the proton mass and $n$ the number density), proton velocity $\bup$, proton 
parallel and perpendicular pressures $p_{\parallel p}$, $p_{\perp p}$ and heat fluxes $q_{\parallel p}$, $q_{\perp p}$, 
together with the induction equation for magnetic field $\bb$. The equations are normalized and density,
magnetic field and proton velocity are measured in units of equilibrium density $\rho_0$, ambient magnetic field $B_0$, 
and Alfv\'en speed $V_A=B_0/\sqrt{4\pi\rho_0}$, respectively. Pressures are measured in units of initial proton parallel
pressure $p_{\parallel p}^{(0)}$ and heat fluxes in units of $p_{\parallel p}^{(0)} V_A$. 
The total pressure in the momentum equation has the form of a tensor. Defining $\bhat=\bb/|\bb|$ as a 
unit vector in the direction of local magnetic field, the proton pressure can be cast in 
the form $\boldsymbol{p}_p=p_{\perp p} \boldsymbol{n} + p_{\parallel p} \boldsymbol{\tau} + \boldsymbol{\Pi}$, where
$\boldsymbol{\tau}=\bhat\otimes\bhat$, and $\boldsymbol{n}=\boldsymbol{I}-\bhat\otimes\bhat$, 
with $\boldsymbol{I}$ being the unit tensor. Finite Larmor radius corrections to the gyrotropic pressures 
are represented by $\boldsymbol{\Pi}$. 
Operator $\otimes$ represents the usual tensor product and in the index notation, for example, $\tau_{ij}=\hat{b}_{i}\hat{b}_{j}$. 
Electrons are assumed to be isothermal with the scalar pressure $p_e=n T_e^{(0)}$, 
where $T_e^{(0)}$ is the electron temperature. Parameter $R_i$, whose inverse multiplies  
the Hall-term and also the FLR corrections, is defined as $R_i=L/d_i$, where $d_i$ 
is the ion inertial length and $L$ is the unit length. The proton plasma beta is defined with 
respect to parallel pressure and $\beta_\parallel=8\pi p_{\parallel p}^{(0)}/B_0^2$.
The density, momentum and induction equations of the FLR-Landau fluid model can then be expressed as    
\begin{eqnarray}
&& \frac{\pr \rho}{\pr t} + \bnabla\cdot(\rho\bup ) = 0, \label{eq:cont}\\ 
&& \frac{\pr \bup}{\pr t} + \bup\cdot\bnabla\bup + \frac{\beta_\parallel}{2\rho} \bnabla\cdot 
(p_{\perp p} \boldsymbol{n} + p_{\parallel p} \boldsymbol{\tau} + \boldsymbol{\Pi} + p_e \boldsymbol{I} ) \nonumber \\
&& -\frac{1}{\rho} (\bnabla\times\bb)\times\bb = 0, \label{eq:mom} \\
&& \frac{\pr\bb}{\pr t} = \bnabla\times(\bup\times\bb) - \frac{1}{R_i}\bnabla\times 
\left[ \frac{1}{\rho} (\bnabla\times\bb)\times\bb   \right]. \label{eq:ind}   
\end{eqnarray}
Dropping, for simplicity, indices $p$ for proton velocity $\bup$ and proton pressures $p_{\perp p},p_{\parallel p}$, 
the evolution equations for perpendicular and parallel pressures reads 
(neglecting the work done by the FLR stress forces) 
\begin{eqnarray}
&& \frac{\pr p_\perp}{\pr t} + \bnabla\cdot(p_\perp \bu) +p_\perp \bnabla\cdot\bu 
-p_\perp \bhat\cdot(\bnabla\bu)\cdot\bhat \nonumber\\
&& + \bnabla\cdot(q_\perp \bhat) + q_\perp \bnabla\cdot\bhat = 0, \label{eq:pperpm}\\
&& \frac{\pr p_\parallel}{\pr t} + \bnabla\cdot(p_\parallel \bu) 
+ 2 p_\parallel \bhat\cdot(\bnabla\bu)\cdot\bhat +\bnabla\cdot(q_\parallel \bhat) 
-2q_\perp \bnabla\cdot\bhat  = 0. \label{eq:pparallelm}
\end{eqnarray}
Assuming an ambient magnetic field of amplitude $B_0$ in the positive z-direction, 
a semi-linear description of the finite Larmor radius corrections in the pressure tensor 
neglecting heat flux contributions can be expressed as 
\begin{eqnarray}
&& \Pi_{xx} = -\Pi_{yy} = -\frac{\langle p_\perp \rangle}{2R_i} (\pr_y u_x +\pr_x u_y), \label{eq:FLRfirst}\\
&& \Pi_{xy} = \Pi_{yx} = -\frac{\langle p_\perp \rangle}{2R_i}(\pr_y u_y - \pr_x u_x), \\
&& \Pi_{yz} = \Pi_{zy} = \frac{1}{R_i} \left[ 2\langle p_\parallel \rangle \pr_z u_x 
+ \langle p_\perp \rangle (\pr_x u_z -\pr_z u_x) \right], \\
&& \Pi_{xz} = \Pi_{zx} = -\frac{1}{R_i}\left[ 2\langle p_\parallel \rangle \pr_z u_y
 + \langle p_\perp \rangle (\pr_y u_z -\pr_z u_y ) \right],\\
&& \Pi_{zz} = 0 \label{eq:FLRlast}, 
\end{eqnarray}
where $\langle p_\perp \rangle$ and $\langle p_\parallel \rangle$ represents the instantaneous averaged 
ion pressures over the entire domain,
whose time variation is aimed to take into account the evolution of the global properties of the plasma.
Finally, parallel and perpendicular heat fluxes $q_\parallel$, $q_\perp$ evolve according to 
\begin{eqnarray}
&& \left( \frac{d}{dt} + \frac{\sqrt{\pi \beta_\parallel}}{4(1-\frac{3\pi}{8})} \mathcal{H}
\pr_z \right) q_\parallel = \frac{1}{1-\frac{3\pi}{8}} \frac{\beta_\parallel}{2}
\pr_z (p_\parallel -\rho), \label{eq:qpar}\\
&& \left( \frac{d}{dt} - \frac{\sqrt{\pi\beta_\parallel}}{2} \mathcal{H} \pr_z \right) 
q_\perp = \nonumber \\
&& \frac{\beta_\parallel}{2} \frac{T_{\perp p}^{(0)}}{T_{\parallel p}^{(0)}} \pr_z
\left[ \left( 1-\frac{T_{\perp p}^{(0)}}{T_{\parallel p}^{(0)}} \right) |b|
- \left( \frac{T_{\parallel p}^{(0)}}{T_{\perp p}^{(0)}} p_\perp - \rho\right) \right], \label{eq:qperp} 
\end{eqnarray}
where $T_{\perp p}^{(0)}, T_{\parallel p}^{(0)}$ are the initial perpendicular and parallel proton temperatures
and $d/dt$ is the convective derivative.
The operator $\mathcal{H}$, which is defined as
\begin{equation}
\mathcal{H} f(z) = - \frac{1}{\pi} VP \int_{-\infty}^{+\infty} \frac{f(z')}{z-z'} dz',
\end{equation}
reduces in the Fourier space to a simple multiplication by $ik_z/|k_z|$ and, is the signature of 
the linear Landau damping.

To gain physical insight into a quite complicated model (\ref{eq:cont})-(\ref{eq:qperp}), it is useful to 
momentarily consider just the largest scales by putting $1/R_i\rightarrow 0$, which 
eliminates the nongyrotropic contributions to the pressure tensor and which 
also eliminates the Hall term. The resulting set of equations still contains the linear Landau damping 
and with the exception of isothermal electrons, it is analogous to the model of 
Snyder, Hammett and Dorland \cite{Snyder1997}. If this model is further
simplified by elimination of eq. (\ref{eq:qpar}), (\ref{eq:qperp}) 
and by instead prescribing $q_\parallel=q_\perp=0$, the resulting model collapses to the
double adiabatic model (Chew et al. \cite{Chew1956}, see also Kulsrud \cite{Kulsrud1983}) and the Landau damping disappears. 
The presence of ion Landau damping in the system (\ref{eq:cont})-(\ref{eq:qperp}) 
is therefore a result of closure equations (\ref{eq:qpar}), (\ref{eq:qperp}) for the heat fluxes, 
which contain the operator ${\mathcal H}$. 
At least in the static limit, this closure can be viewed as a modified Fick's law where the gradient operator 
that usually relate the heat flux to the temperature fluctuations is here replaced by a Hilbert transform that is a 
reminiscence of the plasma dispersion function arising in the linear kinetic theory, and is a signature of Landau 
(zero-frequency) wave-particle resonance.
The effect of Landau damping in the system 
(\ref{eq:cont})-(\ref{eq:qperp}) might be better understood by solving the linearized set 
of equations. In the Appendix we consider waves which 
propagate parallel to the ambient magnetic field. It is shown that except the Alfv\'en waves,  
linear waves have a frequency with a negative imaginary part, and are therefore damped. 
This corresponds to linear Landau damping.
 
The model used for the simulations presented here has several limitations. 
First of all, it contains electrons which are assumed isothermal, a regime in fact often assumed 
in hybrid simulations which provide a kinetic description of the ions and a fluid description of the 
electrons. In the future, more realistic simulations will be performed with inclusion of independent evolution equations
for parallel and perpendicular electron pressures and heat fluxes. This will also result 
in the presence of electron Landau damping, which is absent in the model presented here and 
which seems to play an important role in solar wind turbulence. Another main limitation
appears to be the form of finite Larmor radius corrections, which are derived as 
a large-scale limit of FLR corrections of the full Landau fluid model and which are 
therefore significantly simplified. The FLR corrections (\ref{eq:FLRfirst})-(\ref{eq:FLRlast}) are 
sufficient for the simulations of freely-decaying turbulence, which do not lead to
significant temperature anisotropies. However, our preliminary simulations which employ 
forcing and lead to strong temperature anisotropies show 
that the FLR corrections (\ref{eq:FLRfirst})-(\ref{eq:FLRlast}) are overly simplified and might lead to 
artificial numerical instabilities. For simulations with 
strong temperature anisotropies, a more refined description of FLR corrections is 
required (see Passot and Sulem \cite{PassotSulem2007}, Borgogno et al. \cite{Borgogno2007}). 


To explore the behavior of the FLR-Landau fluid model, we performed simulations of freely decaying turbulence.
The code we used is based on a pseudo-spectral discretization method, where spatial derivatives are evaluated in Fourier  
space. The time stepping is performed in real space with a 3rd order Runge-Kutta scheme. Spatial resolution is $N^3 =128^3$
and the size of the simulation domain is $L=16\times(2\pi)$ in each direction.  
The Hall parameter is $R_i=1$, implying that the lengths are measured in the units of ion inertial length $d_i$. 
Velocity and magnetic field fluctuations of mean square root amplitude 
$\langle \bu^2 \rangle ^{1/2} = \langle \bb^2 \rangle ^{1/2} = 1/8$ 
are initialized in Fourier space 
on the first four modes $k d_i=m/16$, where $m\in [1,4]$, 
with flat spectra and random phases and are constructed to be divergence free. 
Constant pressures $p_\perp=p_\parallel=1$, density $\rho=1$ and heat fluxes $q_\perp=q_\parallel=0$ are initialized in
the entire domain. The temperature of the electrons is $T_e^{(0)}=1$ and is therefore equal to proton temperatures, 
which can be defined as $T_\perp = \langle p_\perp/\rho\rangle$ and $T_\parallel = \langle p_\parallel/\rho\rangle$.
In the simulations presented here, both $T_\perp$ and $T_\parallel$ stay rather close to their initial value.
The compressible Hall-MHD model consists of eq. (\ref{eq:cont})-(\ref{eq:ind}), 
where the divergence of the pressure tensor in eq. (\ref{eq:mom}) is substituted 
with the usual gradient of scalar pressure and, assuming 
the adiabatic law $p=\rho^\gamma$, in the normalized units it is equal 
to $\beta_0/\gamma\bnabla \rho^\gamma$, where $\beta_0$ is the usual
plasma beta defined as $\beta_0=c_s^2/V_A^2$ and the sound speed $c_s^2=\gamma p/\rho$. Adiabatic index $\gamma=1.66$
is used. 

As shown for example by Hirose et al. \cite{Hirose2004} and Howes \cite{Howes2009}, it is not straightforward to compare
Hall-MHD and Vlasov-Maxwell kinetic theory because the associated dispersion relations are quite different if
in the kinetic description the proton (ion) temperature $T_p$ is not negligible with respect to the electron temperature $T_e$. 
We choose to follow Howes \cite{Howes2009}, who, in order to compare Hall-MHD with the kinetic theory 
(which is represented here by the FLR-Landau fluid model), 
defined the necessary relation between $\beta_0$ and $\beta_\parallel$ as $\beta_0=\beta_\parallel(1+T_p/T_e)/(2T_p/T_e)$. 
For equal proton and electron temperatures $T_p=T_e$ this relation yields $\beta_0=\beta_\parallel$.
Landau damping alone is not sufficient
to run the code and some kind of artificial dissipation is needed to terminate the cascade. 
We here resorted to use a filtering in the form of $\frac{1}{2}\{1-\tanh[ (m-0.8N/4)/3]\}$, 
where $m$ is the mode index, applied
each time step on all fields for both Hall-MHD and Landau fluid regimes. Because of the filtering, it 
is crucial to have identical time steps $dt=0.128$ in both models. 
In most of the simulations we used $\beta_0=\beta_\parallel=0.8$. 
In normalized units, the Alfv\'en speed is equal to unity, the usual MHD sound speed $c_s=\sqrt{\beta_0}=0.894$
and the turbulent sonic Mach number $M_s=\langle \bu^2 \rangle ^{1/2}/c_s=0.14$. In section IV., we also 
consider simulations with $\beta_0=\beta_\parallel=0.25$ ($M_s=0.25$) and $\beta_0=\beta_\parallel=0.1$ ($M_s=0.4$).

\section{Identification of the MHD modes}
A wave analysis procedure was implemented in the code, which consists in choosing few 
modes in spatial Fourier space for each field and recording their value every 20 time steps. After the run, 
the time Fourier transform is performed and frequency-power spectra are obtained for each mode. 
This procedure makes it possible to identify at which frequency there is maximum power for each mode, 
and by comparing these with frequencies obtained from theoretical dispersion relations it allows 
to uniquely identify which waves are present in the system. This method was previously used 
for incompressible MHD by Dmitruk and Matthaeus \cite{DmitrukMatthaeus2009}, therefore detecting Alfv\'en waves. 
Considering the dynamics associated with waves propagating in the direction parallel to the ambient magnetic field 
(propagation angle $\theta=0^\circ$), Fig. 1 shows frequency-power spectra of four modes with wavenumbers 
$k_x=0, k_y=0, k_z d_i=m/16$, with $m=1,2,4,8$, 
recorded from the component $b_x$ for Hall-MHD (left) and for FLR-Landau fluid (right). 
\begin{figure}
$$\includegraphics[width=0.5\textwidth]{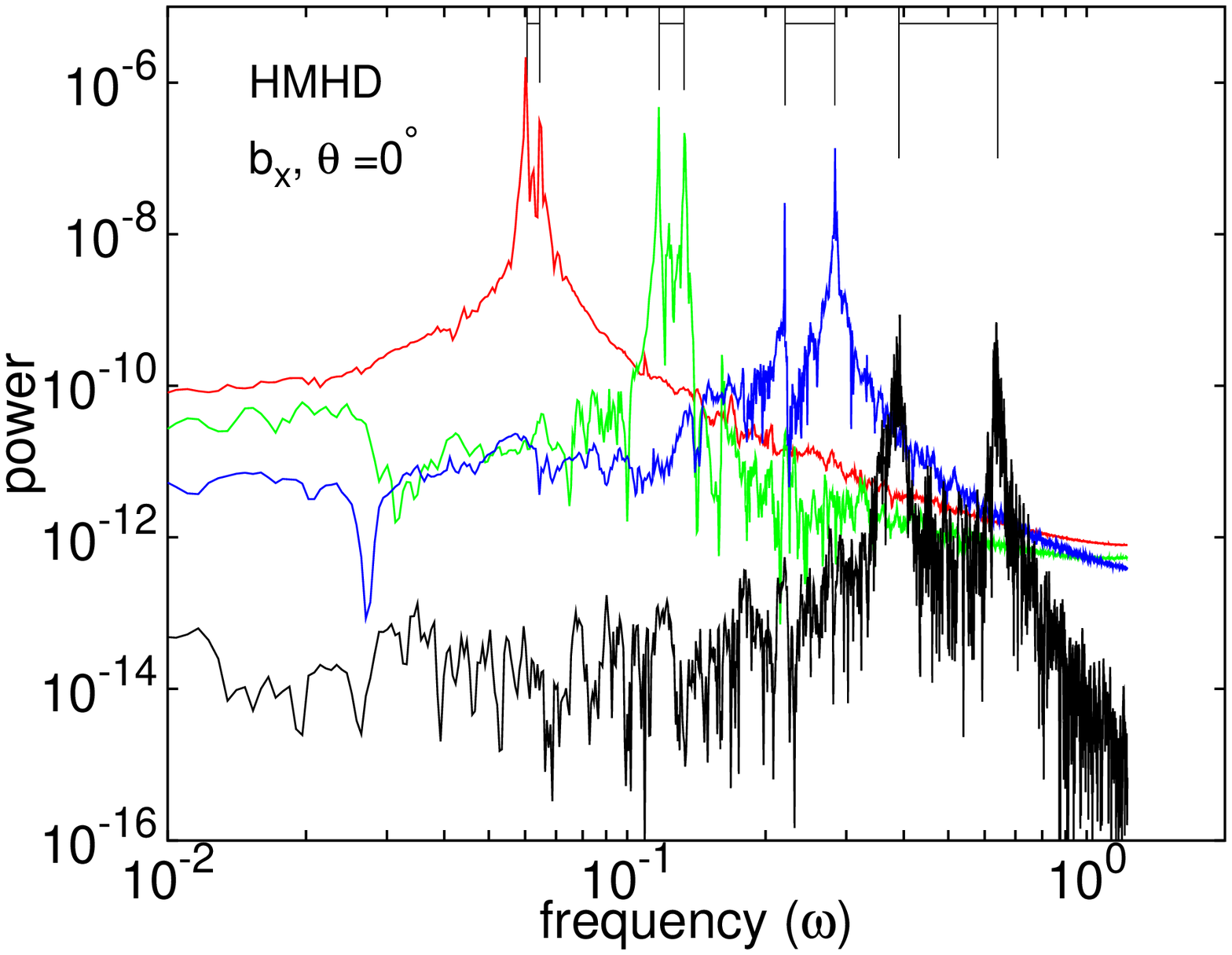}
\includegraphics[width=0.5\textwidth]{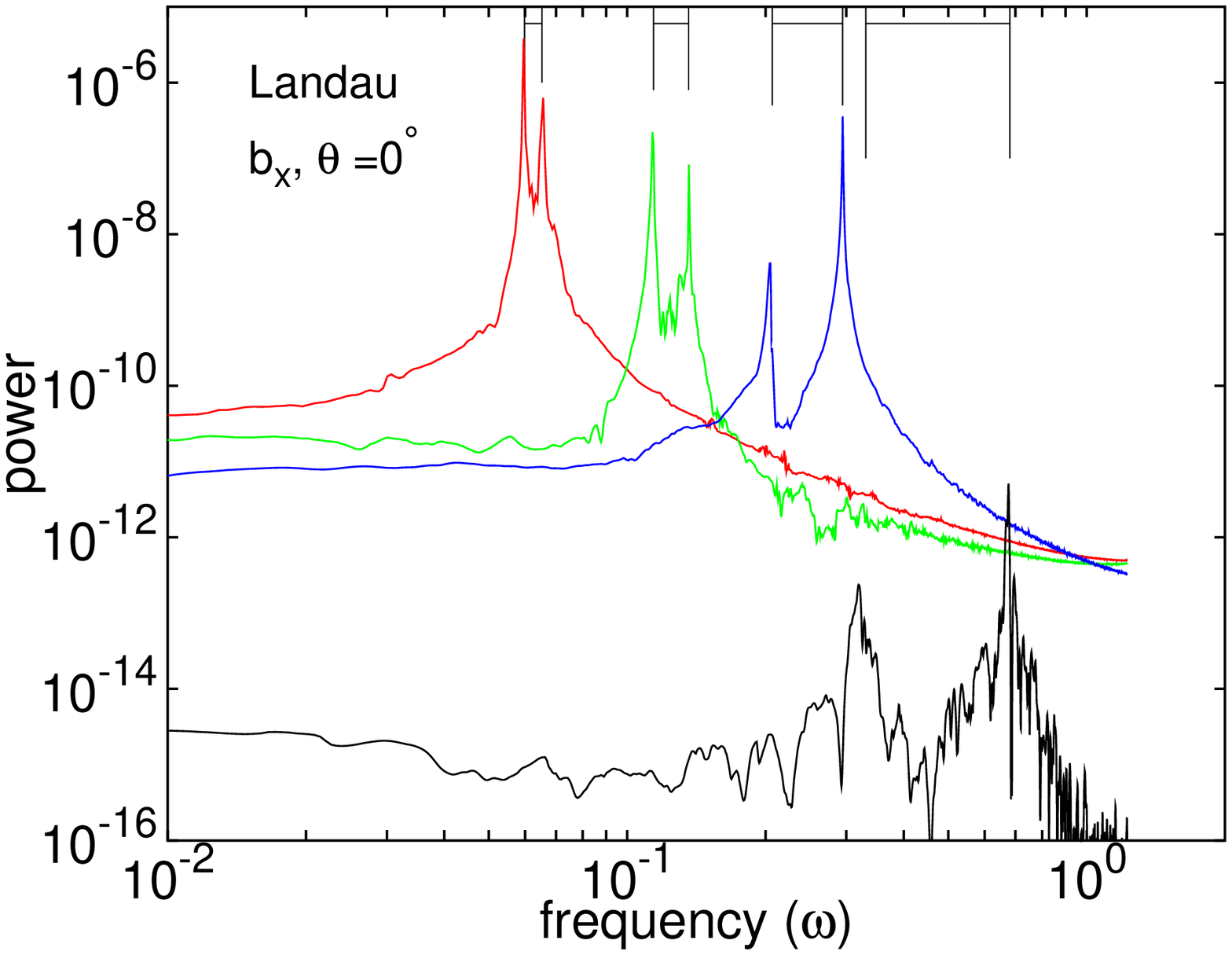}$$
\caption{Left and right polarized Alfv\'en waves for the propagation angle of $0^\circ$ for Hall-MHD (left) 
and Landau fluid (right), as revealed by frequency analysis of 
$b_x$ modes with wavenumbers $k_x=0, k_y=0, k_z d_i=m/16$, where $m=1$ (red), 
$m=2$ (green), $m=4$ (blue), $m=8$ (black). 
Theoretical predictions for peaks calculated from dispersion relations for given $k$ are shown on the top axis.}
\end{figure} 
For Hall-MHD, 
these correspond to polarized Alfv\'en waves which obey the dispersion relation 
$\omega = \pm k^2/(2R_i) + k\sqrt{1+(k/2R_i)^2}$. The theoretical frequency values which are expected from this 
dispersion relation for given $k$ are plotted on the top axis of the figure as black vertical lines.
Figure 1 shows that the resolution of $128^3$ is sufficient to clearly distinguish between left and right 
polarized Alfv\'en waves. In contrast with Hall-MHD, for which it is possible to write the general 
dispersion relation for frequency $\omega$ as a relatively simple polynomial of 6th order, 
for FLR-Landau fluids the general dispersion relation would be uneconomically large to write down.  
In general, it is necessary to numerically solve the determinant obtained 
from linearized equations (\ref{eq:cont})-(\ref{eq:qperp}) for a given wavenumber $k$ after assuming linear waves.
FLR-Landau fluid (\ref{eq:cont})-(\ref{eq:qperp}) consists of 11 evolution equations in 11 variables and, 
together with the divergence free constraint for the magnetic field, therefore yields general dispersion relation for
frequency $\omega$ in the form of a complex polynomial of 10th order. This represents 5 forward and 5 backward 
propagating waves, with some solutions having highly negative imaginary part and which are therefore strongly damped. 
A similar situation is encountered in the Vlasov-Maxwell kinetic theory which essentially yields an infinite number of 
strongly damped solutions. 
For the propagation angle $\theta=0^\circ$ and additional constraint $T_\perp =T_\parallel=1$, it is however 
possible to obtain an analytic solution for the circularly polarized Alfv\'en waves as 
\begin{equation}
\omega = \pm\frac{k^2}{2R_i}\left( 1+\frac{\beta_\parallel}{2} \right) 
+k\sqrt{1+\left( \frac{k}{2R_i}\right)^2 \left( 1-\frac{\beta_\parallel}{2}\right)^2}, \label{eq:zeroAlfven}
\end{equation}
with two other solutions obtained by substituting $\omega$ with $-\omega$. The dispersion relation is 
quite similar to that of Hall-MHD with additional terms proportional to $\beta_\parallel$ and resulting from 
the finite Larmor radius corrections. Expected theoretical frequencies obtained from this analytic solution
are plotted on the top axis of Fig. 1 for the Landau fluid regime (right). They again match quite precisely.  
Note also the moderately strong damping of parallel Alfv\'en waves in Landau fluid regime, which can be seen
for the last mode $m=8$. Landau damping does not act directly on linear constant amplitude Alfv\'en waves 
obeying relation (\ref{eq:zeroAlfven}), which are exact solutions of linearized FLR-Landau fluid model. 
However, nonlinear parallel Alfv\'en waves, which are of course present
in the full model, cause production of density (sound) fluctuations. Sound waves in the FLR-Landau 
fluid model, as well as in the kinetic theory, are heavily damped by Landau damping as shown below and 
this process therefore also results in damping of Alfv\'en waves. Mj{\o}lhus and Wyller \cite{MjolhusWyller1988}
studied the kinetic derivative nonlinear Schr\"odinger equation (KDLNS) for parallel propagating long-wavelength Alfv\'en 
waves where they refer to this effect as nonlinear Landau damping, because 
it is acting on Alfv\'en waves in a nonlinear way.  

Further exploring propagation angle of $\theta=0^\circ$, Fig. 2 shows frequency power spectra obtained 
from component $u_z$ and which should therefore predominantly
display sound waves (almost identical spectra can be obtained from component $\rho$). 
\begin{figure}
$$\includegraphics[width=0.5\textwidth]{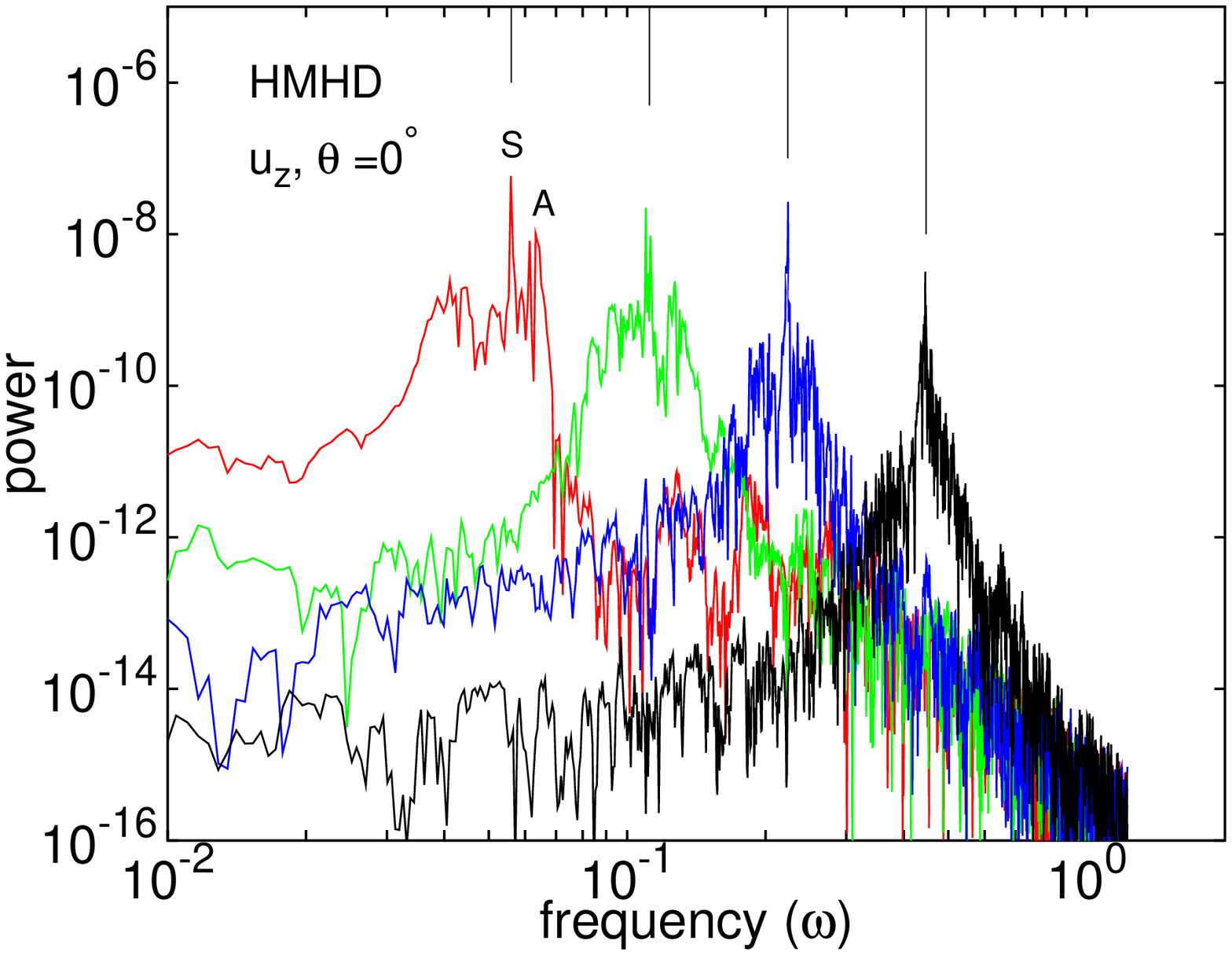}
\includegraphics[width=0.5\textwidth]{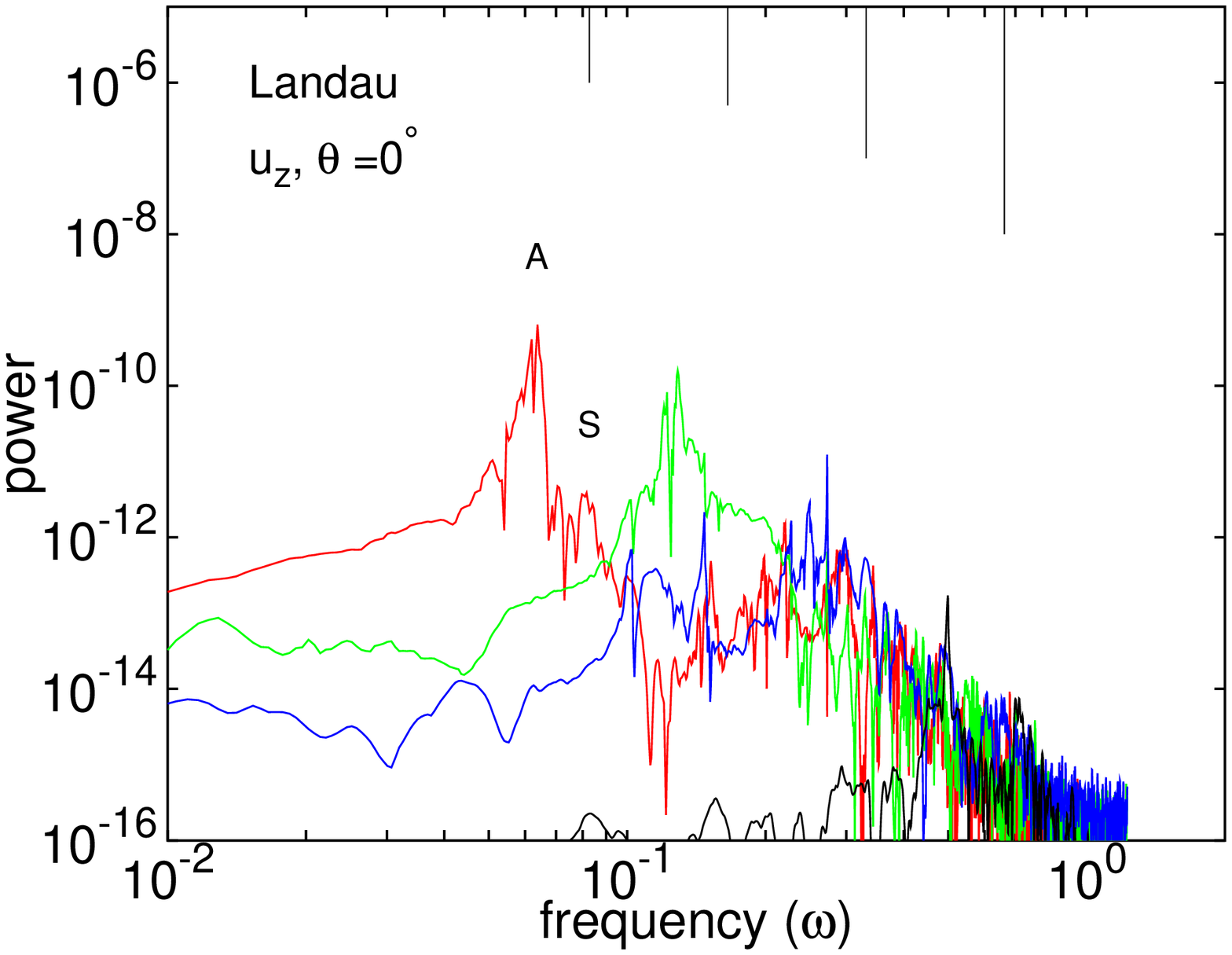}$$
\caption{Sound waves for the propagation angle of $0^\circ$ for Hall-MHD (left) and Landau fluid (right),
from frequency analysis of $u_z$ modes with the same wavenumbers as in Fig. 1. Sound waves (S) are heavily damped for the Landau fluid, 
which is consistent with the kinetic theory. The spectra also show 
weak presence of Alfv\'en waves (A), which are visible for the first two modes.}
\end{figure}
The same modes with $k_z d_i=m/16$, where $m=1,2,4,8$ are shown as in Fig. 1. 
For Hall-MHD, parallel sound waves obey the dispersion relation $\omega=k c_s$, where 
the sound speed $c_s=\sqrt{\beta_0}$. Sound waves (S) are clearly presented in Fig. 2 for Hall-MHD (left) with 
quite sharp peaks which match the theoretical dispersion values shown on the top axis. 
Weak presence of polarized Alfv\'en waves (A) for modes $m=1,2$ is also visible in this component generated by nonlinear coupling. 
For the Landau fluid regime (Fig. 2 right), it is not possible to obtain simple analytic dispersion relation which 
corresponds to sound waves and correct values must be obtained numerically as explained above (see also the Appendix). 
This yields 5 frequencies for forward propagating waves, with 2 solutions corresponding
to the polarized Alfv\'en waves (\ref{eq:zeroAlfven}) and 3 solutions which are highly damped. 
The sound wave was chosen as the least damped solution. Dispersion relation $\omega=k\sqrt{(3+T_e^{(0)})\beta_\parallel/2}$ 
obtained from the double adiabatic model, to which Landau fluid description collapses after assumption of zero heat fluxes
and large scales, can also be used as a heuristic guide to determine which out of the 3 damped solutions 
represents the sound wave frequency.
Obtained frequency values are again plotted on the top axis.
Figure 2 shows that the sound waves (S) are heavily damped for the Landau fluid regime
(Fig. 2 right) with the nonlinearly generated Alfv\'en waves (A) completely overpowering the spectra. 
Inhibition of sound waves is consistent with the kinetic theory. 
As elaborated by Howes \cite{Howes2009}, sound waves are overestimated by  
MHD and Hall-MHD descriptions as they represent solutions which are strongly damped by the Landau resonance.
Note that the calculated sound wave frequencies in the FLR-Landau fluid regime are actually higher 
than the corresponding Alfv\'en wave frequencies. 
A similar result was obtained by Howes \cite{Howes2009}
who numerically compared dispersion relations for Hall-MHD and kinetic theory and who, for the 
nearly parallel propagation with $T_i=T_e$ and $\beta_0=\beta_\parallel=1$, noted that the 
kinetic solution corresponding to the slow wave has a higher phase speed than the kinetic solution
corresponding to the Alfv\'en wave.     
  
Considering the propagation angle of $\theta=45^\circ$, Fig. 3 shows frequency-power spectra in component $b_x$ for 
wavenumbers $k_y=0, k_x d_i=k_z d_i=m/16$, where $m=1,2,4,6$ and which therefore predominantly shows 
slow (S) and fast (F) magnetosonic waves. Spectra also show weaker presence of Alfv\'en waves (A). Again for this angle 
of propagation, the slow waves are strongly damped. In both Hall-MHD and FLR-Landau fluid simulations, the 
associated dispersion relations had to be solved numerically and the predicted frequencies for spectral peaks 
are shown on the top axis. Slow waves (S) in the FLR-Landau fluid regime were again identified as the least damped 
frequency out of the 3 highly damped solutions. For identical wavenumbers, the presence of Alfv\'en waves can be 
better explored in the component $b_y$ and corresponding spectra are shown in Fig. 4.  
\begin{figure}    
$$\includegraphics[width=0.5\textwidth]{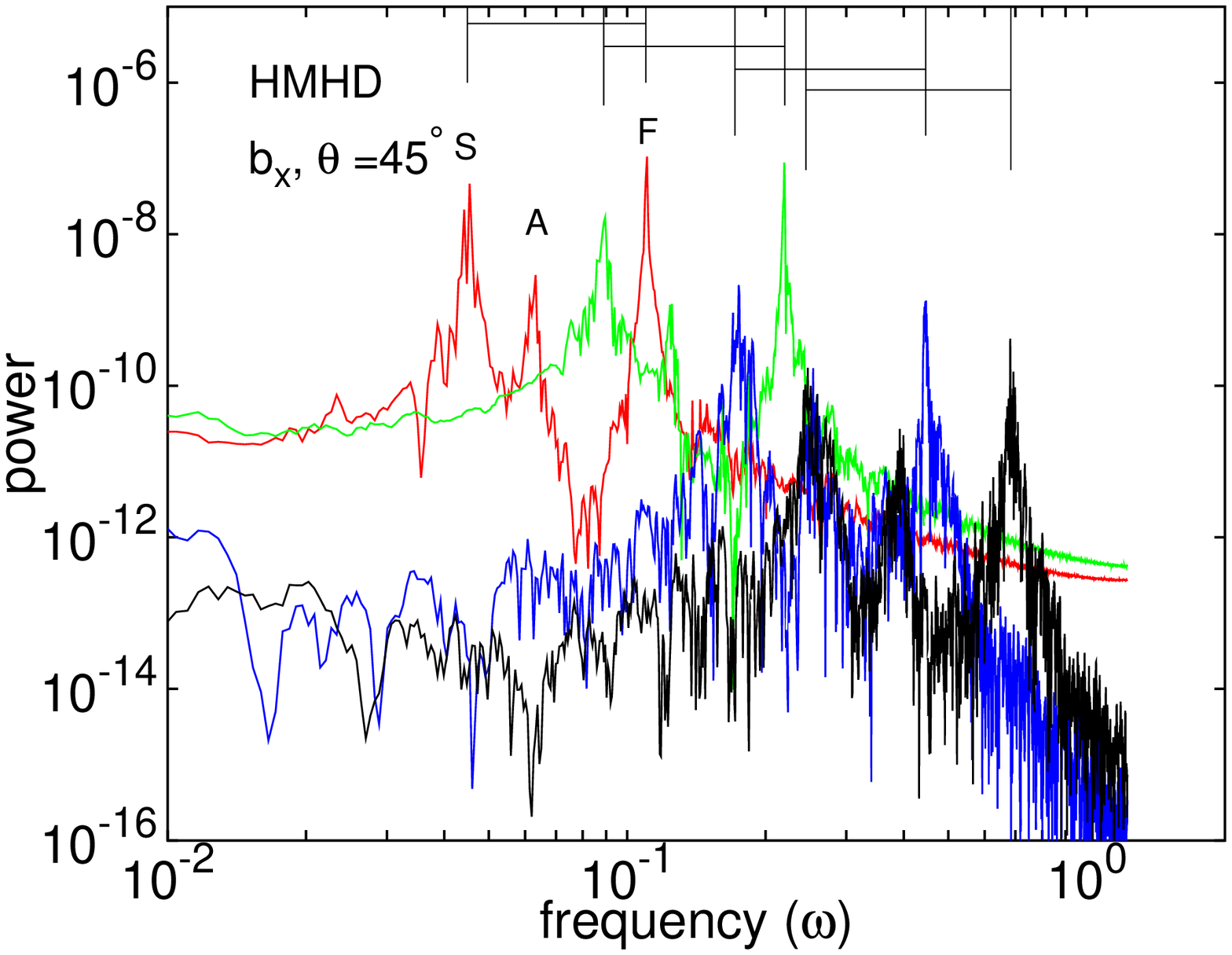}
\includegraphics[width=0.5\textwidth]{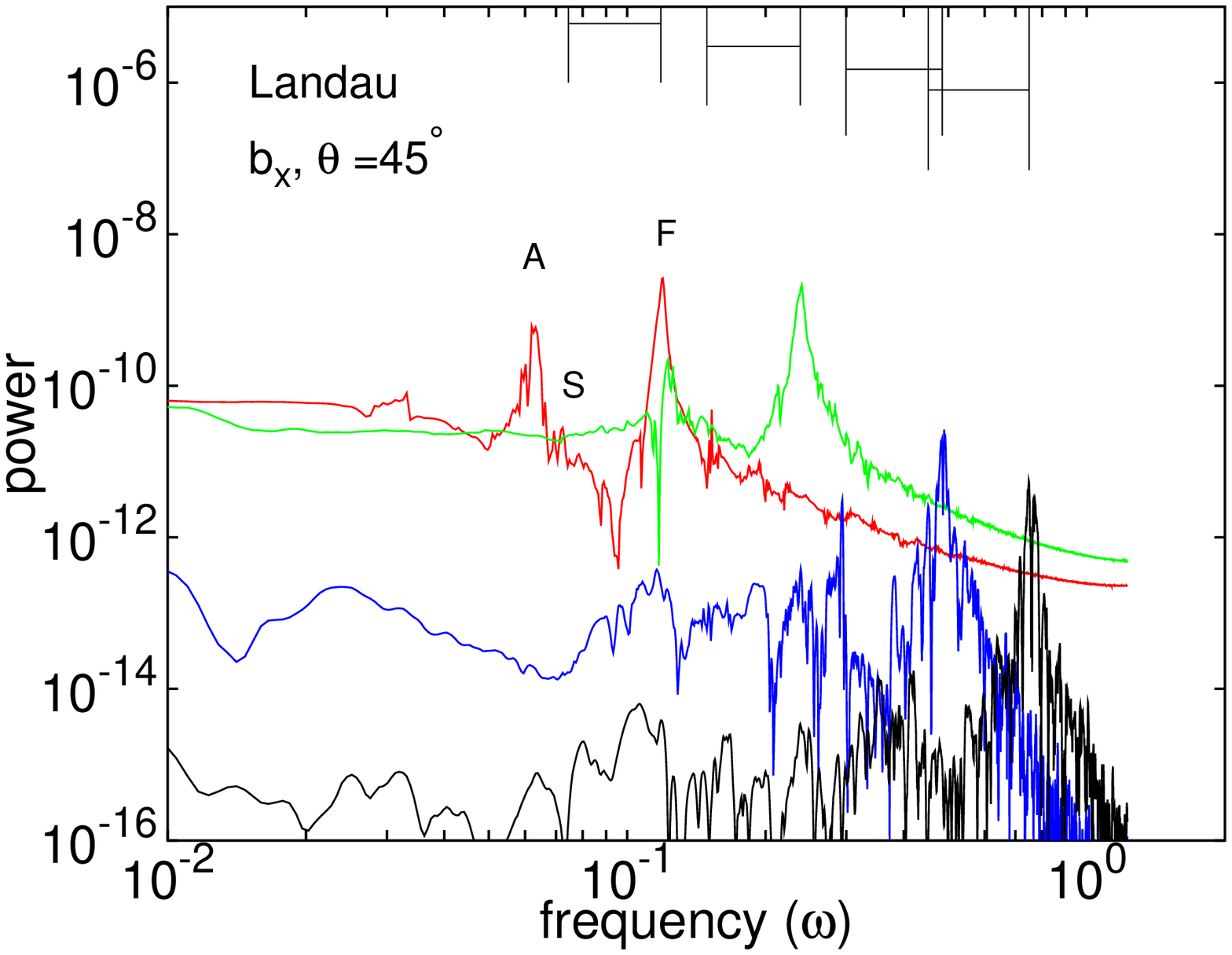}$$
\caption{Slow (S) and fast (F) magnetosonic waves for the propagation angle of $45^\circ$ for 
Hall-MHD (left) and Landau fluid (right), from frequency analysis of $b_x$ modes 
with wavenumbers $k_y=0, k_x d_i=k_z d_i=m/16$, where $m=1$ (red), $m=2$ (green), $m=4$ (blue), 
$m=6$ (black). The presence of Alfv\'en waves (A) is also visible. 
Theoretical predictions for slow and fast waves are shown on the top axis.}
\end{figure}
\begin{figure}
$$\includegraphics[width=0.5\textwidth]{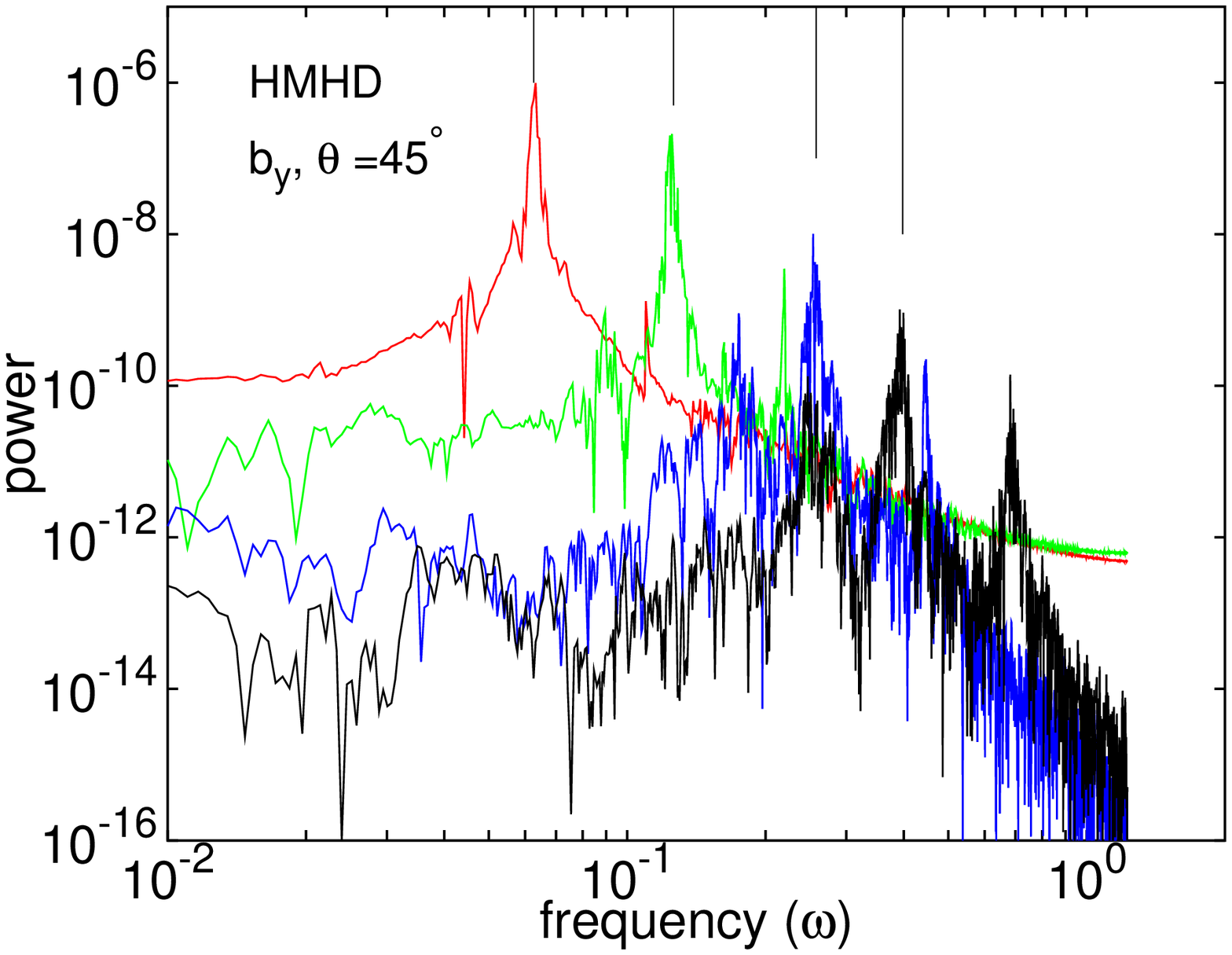}
\includegraphics[width=0.5\textwidth]{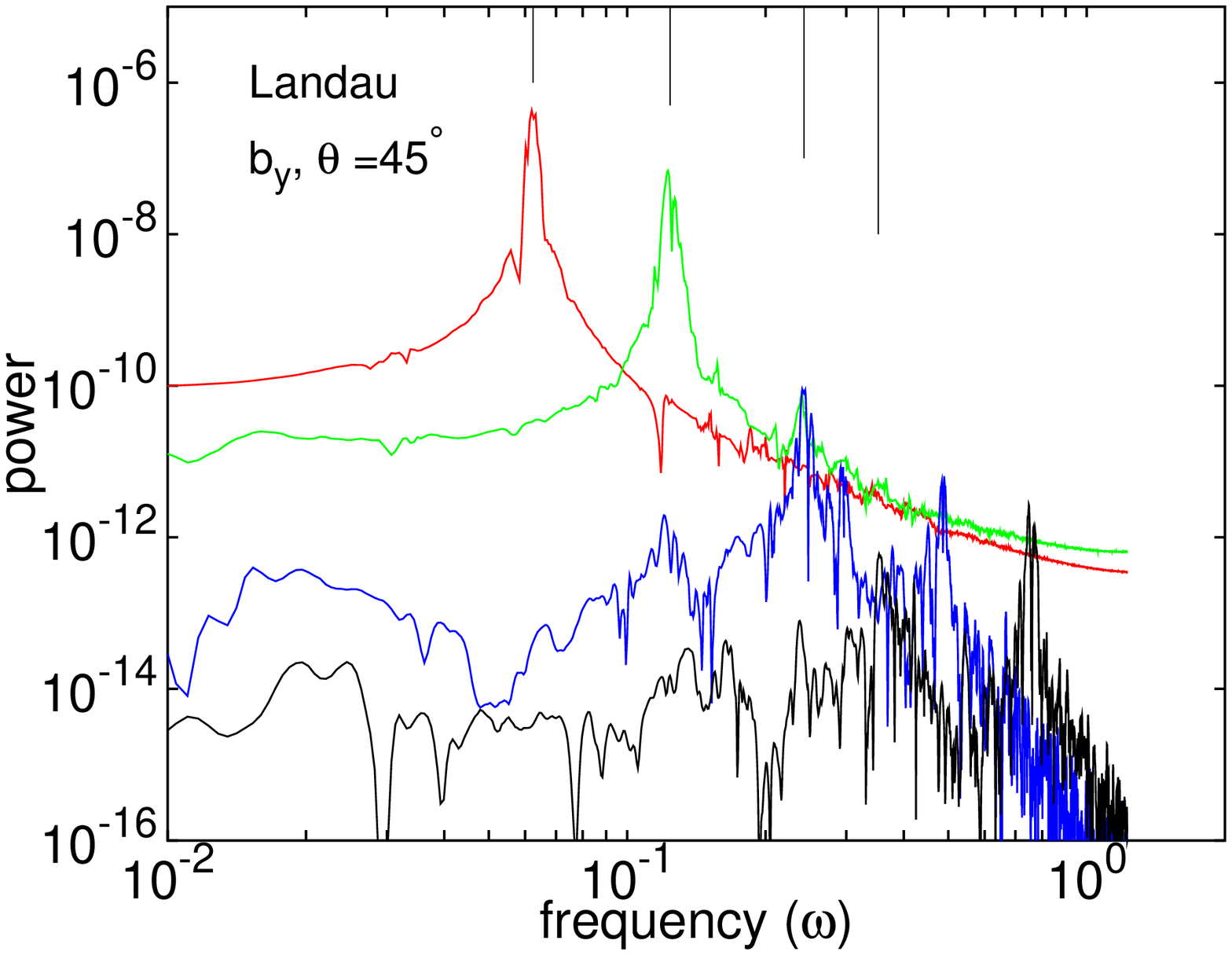}$$
\caption{Alfv\'en waves for the propagation angle of $45^\circ$ for Hall-MHD (left) and Landau fluid (right),
from frequency analysis of $b_y$ modes with the same wavenumbers as in Fig. 3. 
Theoretical predictions for the Alfv\'en waves frequencies are shown on the top axis.}
\end{figure}

Finally, considering purely perpendicular propagation with $\theta=90^\circ$, Fig. 5 shows frequency-power spectra
obtained from the density field $\rho$ and wavenumbers $k_y=0, k_z=0, k_x d_i=m/16$, where $m=1,2,4,8$. For Hall-MHD
regime (Fig. 5 left) the spectral peaks correspond to magnetosonic waves with the usual dispersion relation 
$\omega=k\sqrt{1+\beta_0}$. In the Landau fluid regime, linearized set of equations (\ref{eq:cont})-(\ref{eq:qperp})
with the additional constraint $T_\perp =T_\parallel =1$ 
can be shown to yield the dispersion relation for perpendicular magnetosonic waves in the form
\begin{equation}
\omega = k\sqrt{1+\beta_\parallel \left( 1+\frac{T_e^{(0)}}{2}\right) + \left( \frac{\beta_\parallel k}{4R_i}\right)^2}.
\end{equation}
The dispersion relation clearly shows the effect of inclusion of isothermal electrons 
(for simulations presented here $T_e^{(0)}=1$) 
and also the effect of finite Larmor radius corrections which are represented 
by the last quadratic term. Theoretical predictions
from Hall-MHD and FLR-Landau fluid dispersion relations are again shown on the top axis. 
To clearly show the shift of the peaks
between the two regimes, we also added the theoretical Hall-MHD frequencies to the top 
axis of FLR-Landau fluid regime (Fig. 5 right)
and represent them with the small magenta lines. 
\begin{figure}
$$\includegraphics[width=0.5\textwidth]{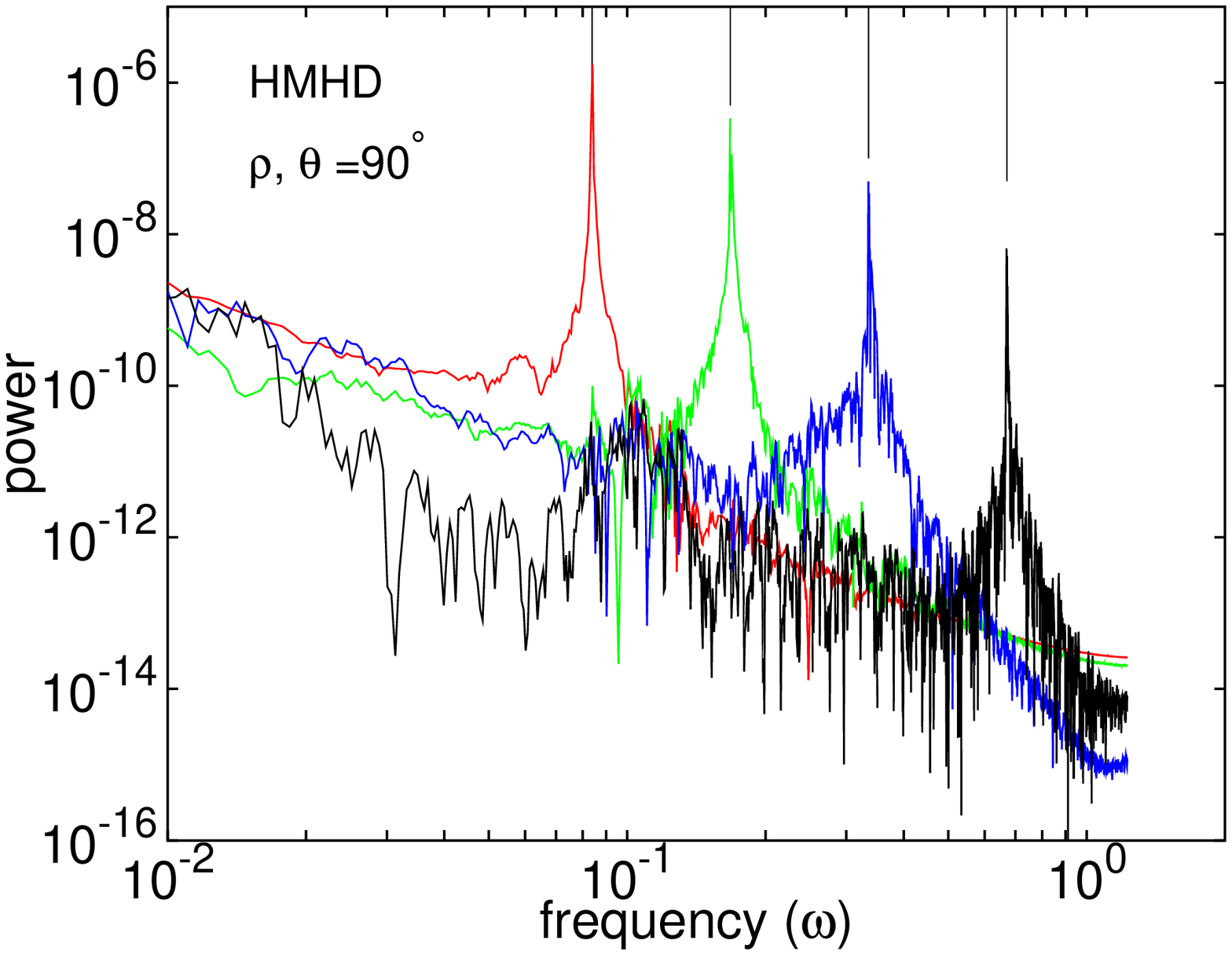}
\includegraphics[width=0.5\textwidth]{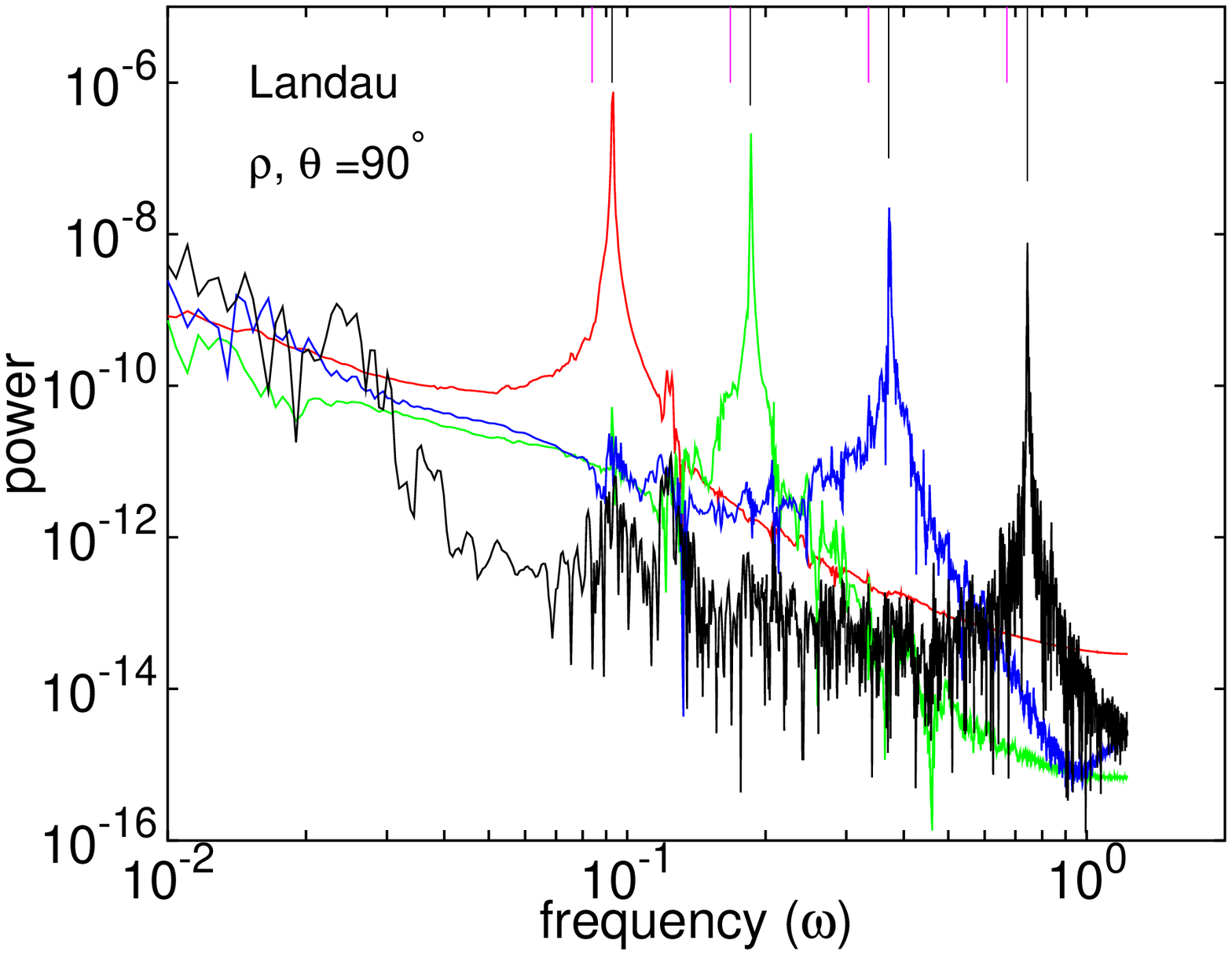}$$
\caption{Magnetosonic waves for the propagation angle of $90^\circ$ for Hall-MHD (left) and Landau fluid (right),
from frequency analysis of density modes 
with wavenumbers $k_y=0, k_z=0, k_x d_i=m/16$, where $m=1$ (red), $m=2$ (green), $m=4$ (blue), $m=8$ (black). 
Theoretical predictions from dispersion relations are shown on the top axis (long black lines). For comparison, on the 
right panel we also included the frequency predictions from Hall-MHD (small magenta lines).}
\end{figure}

\section{Flow compressibility}
Compressibility of the flow can be evaluated by decomposing the velocity field into its solenoidal and 
non-solenoidal components and by calculating the associated energies according to 
\begin{equation} \label{eq:compincomp}
\sum_{\bk} |\buk |^2 = \sum_{\bk} \frac{|\bk\times\buk|^2}{|\bk|^2} + \sum_{\bk} \frac{|\bk\cdot\buk|^2}{|\bk|^2},
\end{equation}
where the left-hand side corresponds to the total energy $E^U$ in velocity 
field, the first term in the right-hand side corresponds to the energy $E_{in}$ in the solenoidal component 
and the second term $E_c$ originates from the compressible one.
Relation (\ref{eq:compincomp}) can be therefore expressed as $E^U=E_{in}+E_c$ and the compressibility of the flow can be 
evaluated as a ratio of compressible and total energy $E_c/E^U$. Time evolution of $E_c/E^U$ for Hall-MHD and FLR-Landau fluid regime 
with $\beta_0=\beta_\parallel=0.8$ is presented in Fig. 6. The figure shows that the ratio $E_c/E^U$ which represents compressibility 
is significantly lower in Landau fluid regime and is therefore a result of presence of Landau damping.
\begin{figure}
$$\includegraphics[width=0.5\textwidth]{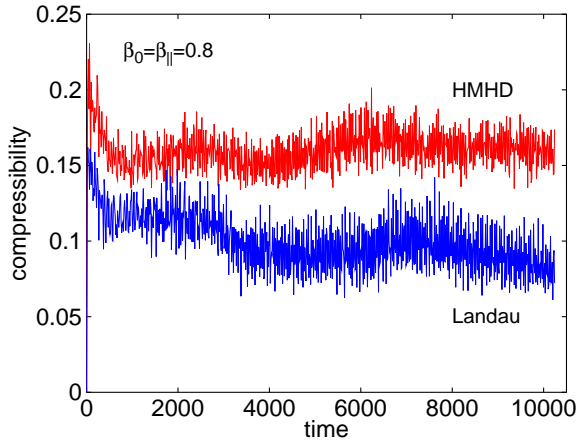}$$
\caption{Compressibility for Hall-MHD (red line) and FLR-Landau fluid (blue line) 
evaluated as $(\sum_k |\bk\cdot\bu_k |^2/|\bk|^2)/\sum_k |\bu_k|^2$ for $\beta_0=\beta_\parallel=0.8$. 
Both regimes start with the identical initial condition where the velocity field is divergence free.
The figure shows that the compressibility is clearly inhibited in the Landau fluid simulation.}
\end{figure}
The question then arises of the influence of the sonic Mach number in the compressibility evolution.
The time evolution of $E_c/E^U$ for $\beta_0=\beta_\parallel=0.25$ (which corresponds to $M_s=0.25$) is displayed 
in Fig. 7 (left), whereas the time evolution of $E_c/E^U$ for $\beta_0=\beta_\parallel=0.1$ 
(which corresponds to $M_s=0.40$) is shown in Fig. 7 (right). The simulations 
were performed with the same time step $dt=0.128$ and filtering, for approximately half the total integration
time of the previous regime with $\beta_0=\beta_{\parallel}=0.8$ ($M_s=0.14$). The figure shows that while the compressibility 
is still clearly reduced in Landau fluid simulation with $\beta_0=\beta_\parallel=0.25$, this reduction 
is almost insignificant for simulations with $\beta_0=\beta_\parallel=0.1$. This is an
expected effect, as the strength of the Landau damping 
is proportional to $\beta_{\parallel}$. We note that the turbulent sonic Mach number in the solar wind 
is typically small and, for example, analysis of observational data 
performed by Bavassano and Bruno \cite{BavassanoBruno1995} (from 0.3-1.0 AU) showed
that the most probable value is between $M_s=0.1-0.2$ with the distribution having an extended tail 
to lower values.  
\begin{figure}
$$\includegraphics[width=0.5\textwidth]{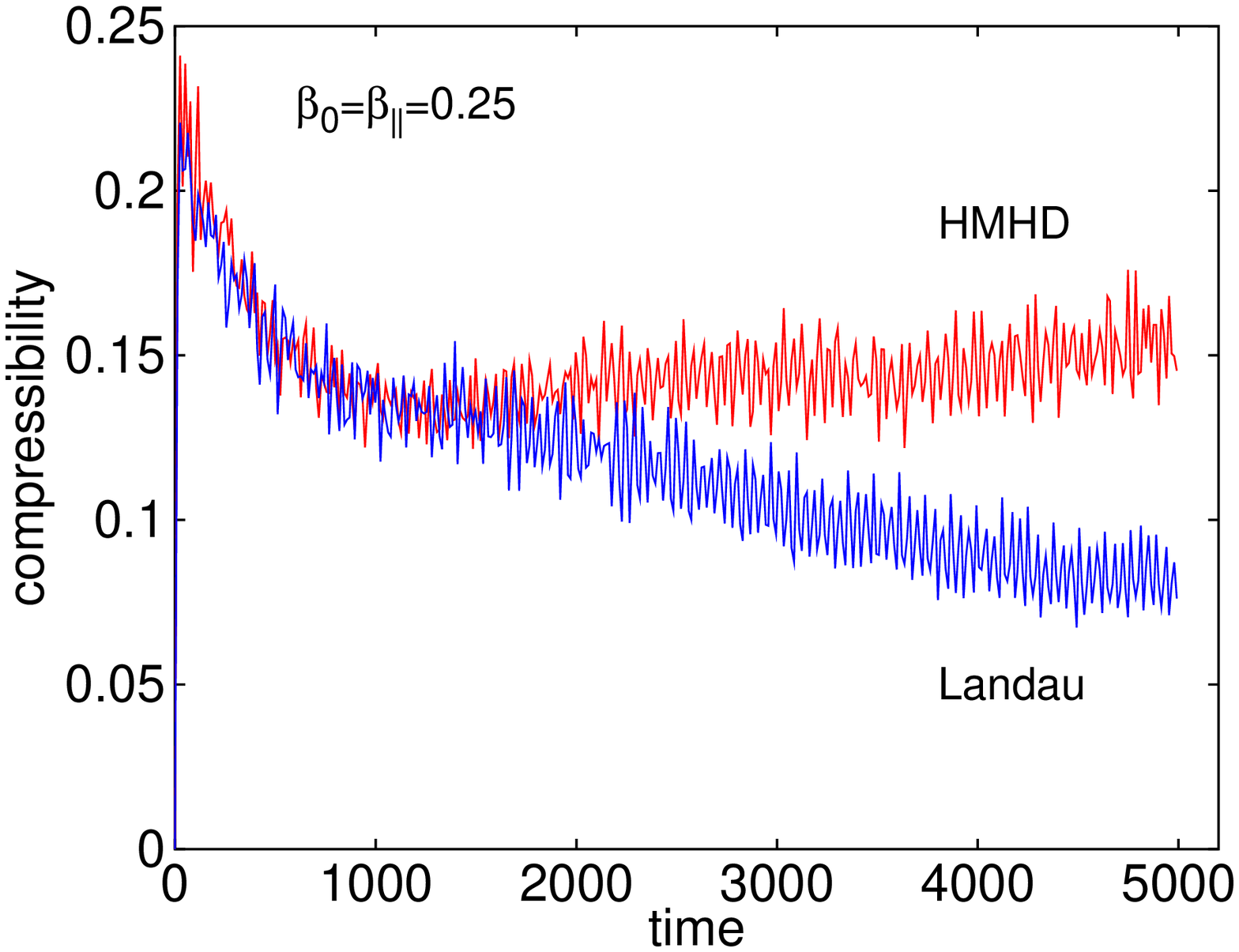}
\includegraphics[width=0.5\textwidth]{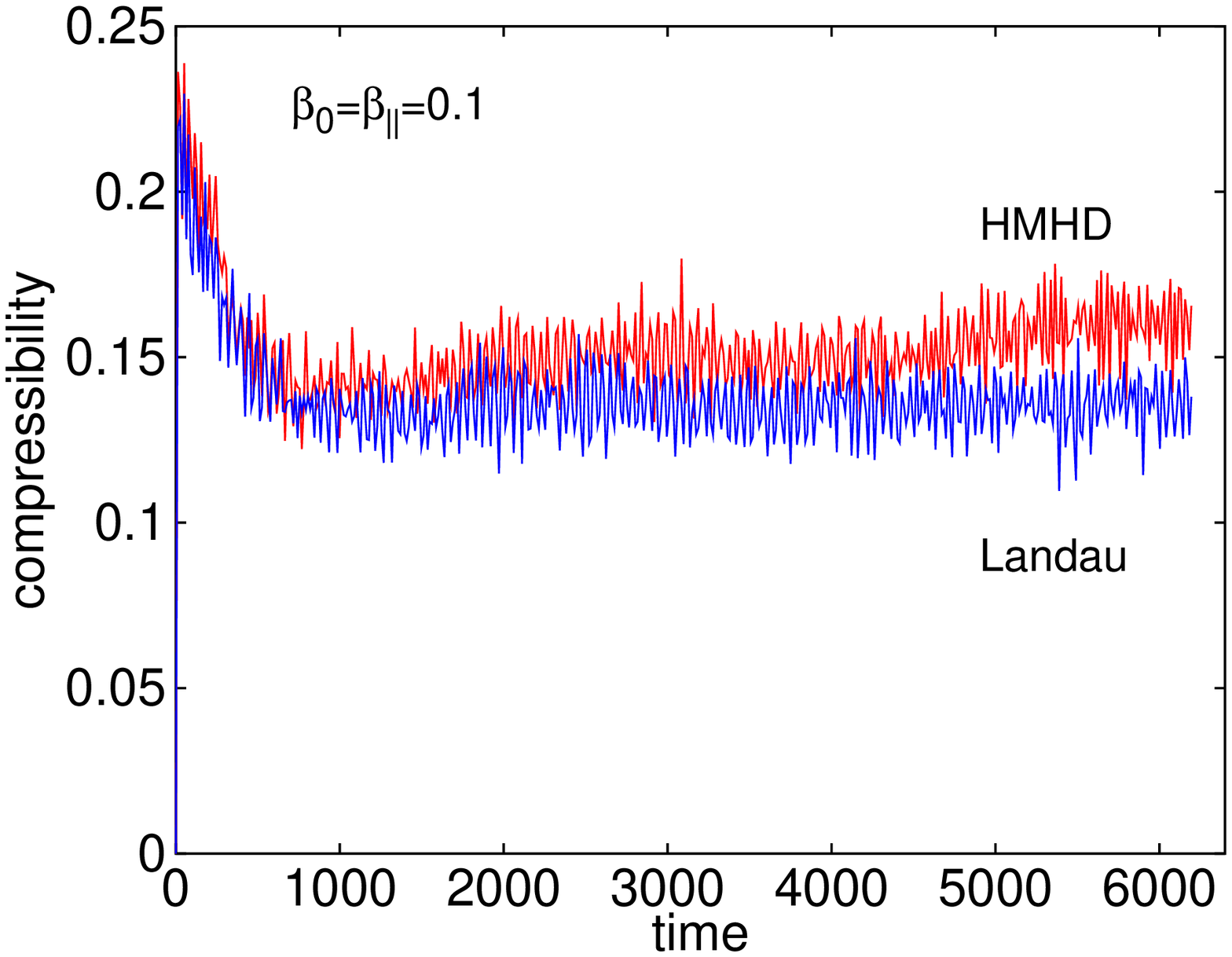}$$
\caption{Compressibility for Hall-MHD (red line) and FLR-Landau fluid (blue line) 
evaluated as $(\sum_k |\bk\cdot\bu_k |^2/|\bk|^2)/\sum_k |\bu_k|^2$ with $\beta_0=\beta_\parallel=0.25$ (left) and 
with $\beta_0=\beta_\parallel=0.1$ (right). The figure shows that the compressibility is clearly inhibited 
in Landau fluid for simulations with $\beta_0=\beta_\parallel=0.25$, whereas for simulations with 
$\beta_0=\beta_\parallel=0.1$ the level of compressibility is almost identical.}
\end{figure}

The question also arises of the compared evolution of total energy for Hall-MHD and FLR-Landau fluid simulations.
The total energy $E_{tot}$ can be evaluated as the sum of the kinetic energy $E_{kin}$, 
the magnetic energy $E_{mag}$ and the internal energy $E_{int}$. 
In Hall-MHD and FLR-Landau fluid model, the definitions of kinetic and  
magnetic energy are identical and equal to
\begin{equation}
E_{kin}=\frac{1}{2}\int \rho |\bu|^2 dx^3, \qquad E_{mag}=\frac{1}{2}\int |\bb|^2 dx^3. \label{eq:Ekin}
\end{equation}
However, the definition of internal energy is naturally different in each model. 
In the Hall-MHD model, the internal energy is defined as
\begin{equation}
\textrm{HMHD:} \qquad E_{int}=\frac{\beta_o}{\gamma(\gamma-1)}\int \rho^\gamma dx^3, 
\end{equation}   
whereas in the FLR-Landau fluid model with isothermal electrons, the internal energy is given by
\begin{equation}
\textrm{Landau:} \qquad E_{int} = \frac{\beta_\parallel}{2}\int \left(p_\perp +\frac{p_\parallel}{2} \label{eq:Eint}
+T_e^{(0)}\rho\ln\rho\right) dx^3.
\end{equation}
It is emphasized that because of the filtering, the total energy $E_{tot}$ is not exactly 
preserved in the Hall-MHD and Landau fluid simulations. During the simulations with $\beta_0=\beta_\parallel=0.8$, 
the total energy in Hall-MHD decreased by approximately 1.2\%, whereas in FLR-Landau fluid the decrease was approximately 0.5\%.
Filtering indeed dissipates kinetic and magnetic energies 
without turning them into heat. Therefore, in contrast with Landau fluid models, 
Hall-MHD simulations do not have heating at all and the internal energy could increase 
only through development of density fluctuations.
\begin{figure}
$$\includegraphics[width=0.5\textwidth]{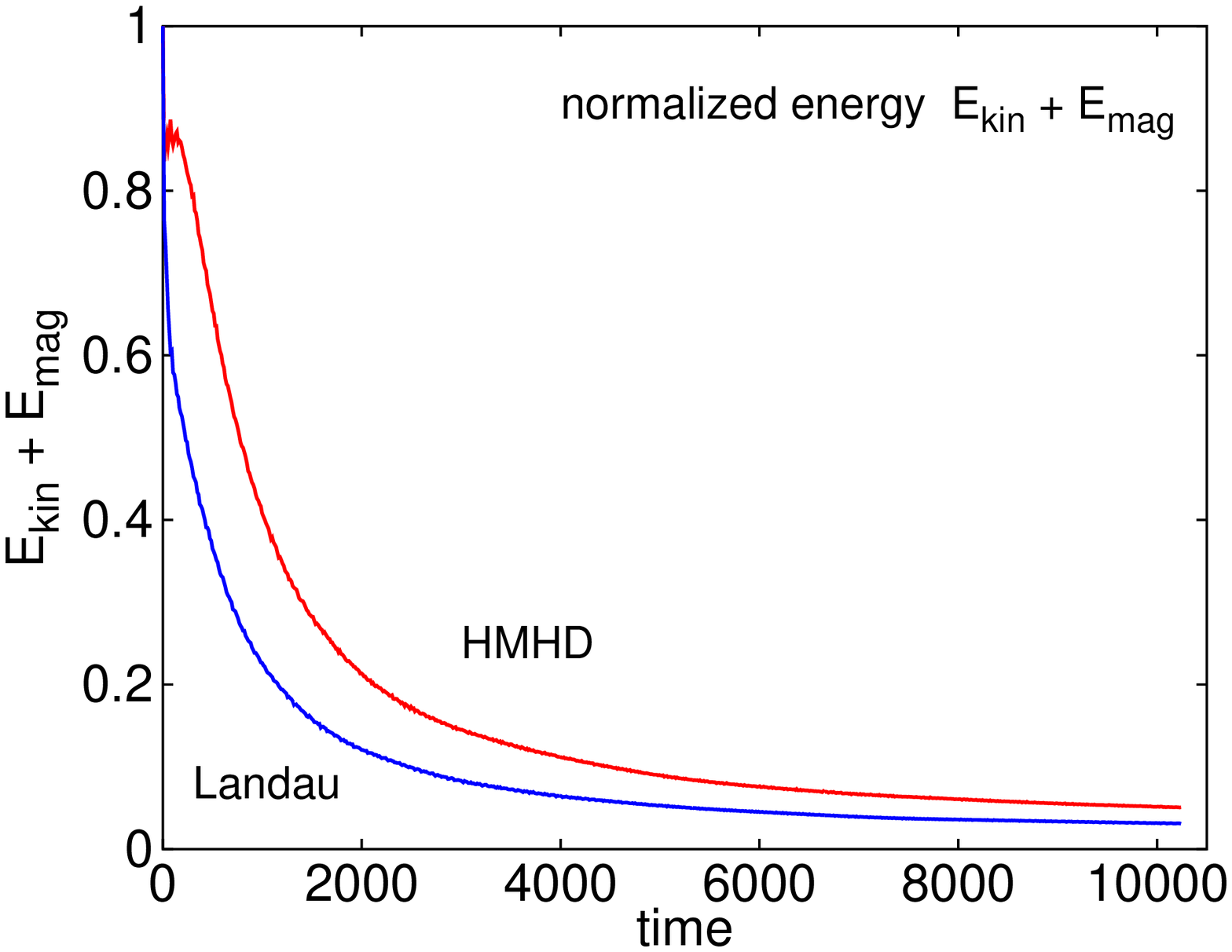}
\includegraphics[width=0.5\textwidth]{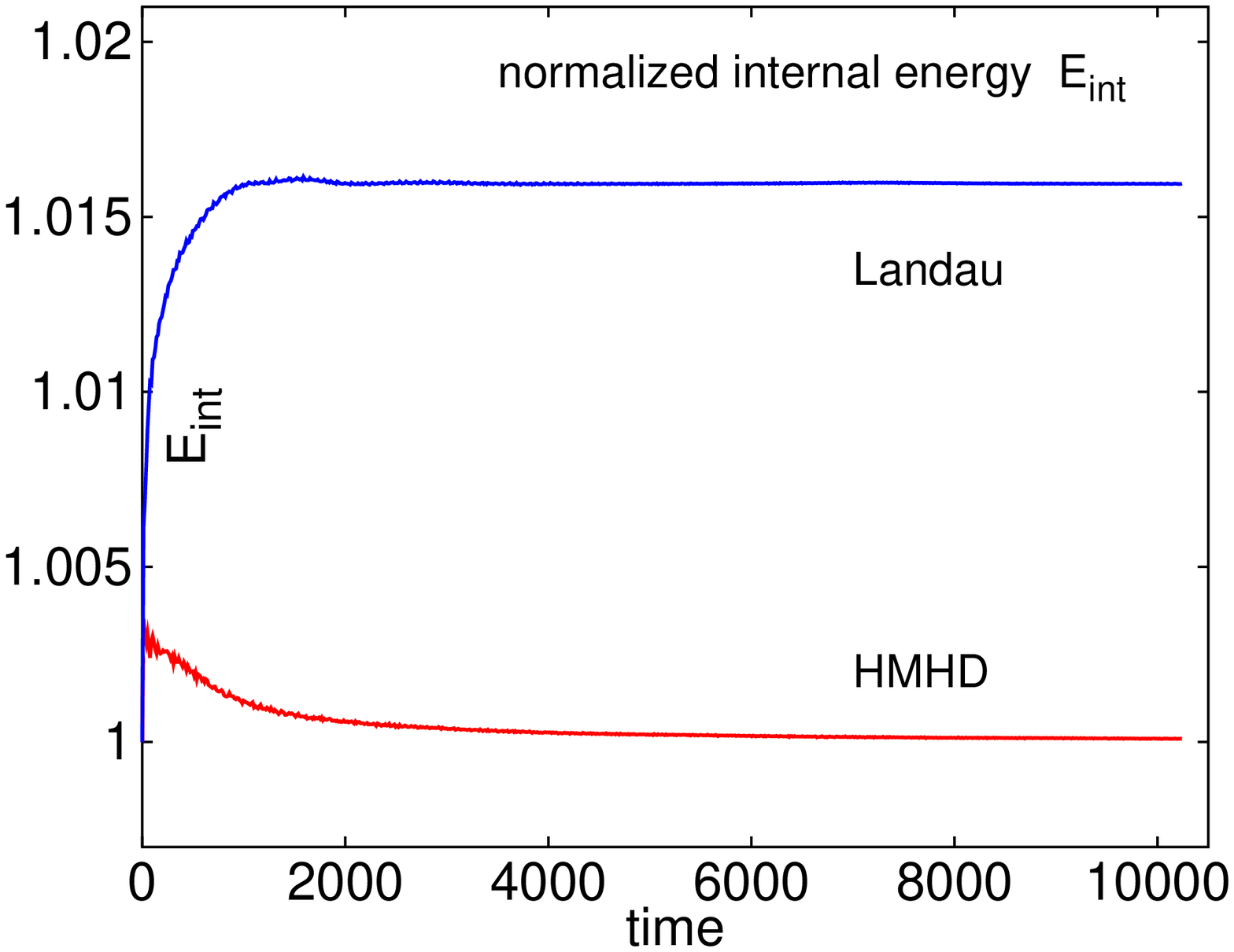}$$
\caption{Normalized mechanical fluctuations energy $E_{kin}+E_{mag}$ (left) and normalized internal energy $E_{int}$ (right) 
for Hall-MHD (red line) and FLR-Landau fluid (blue line), in the case $\beta_0=\beta_\parallel=0.8$. 
Landau damping transfers energy from $E_{kin}+E_{mag}$ into $E_{int}$.}
\end{figure}
Time evolution of the sum of kinetic and magnetic energies (normalized to their initial values)
is displayed in Fig. 8 (left), where the energy contained in the ambient magnetic field was subtracted. This figure shows that this sum  
decays faster for FLR-Landau fluid simulation. Time evolution of internal energy $E_{int}$ 
is shown in Fig. 8 (right), where energies were again normalized to their initial value. 
The initial jump observed in both simulations 
reflects a rapid adjustment from initial conditions that are not close 
to an equilibrium state. Later on, the internal energy of Hall-MHD simulation
decreases, whereas the internal energy of FLR-Landau fluid simulation 
more or less smoothly increases until around time $t = 1500$.
During this time (which corresponds to over $10^4$ time steps) the Landau damping acts strongly  
and, by mainly damping slow waves, converts the mechanical energy $E_{kin}+E_{mag}$ into the internal energy $E_{int}$, 
which represents heating of the plasma.
The question also arises what fraction of mechanical energy is dissipated directly by Landau damping
and what fraction is dissipated by the filtering process. Unfortunately, we are unaware of 
any technique how to address this question.  

We note that for simulations of freely decaying turbulence the 
heating is quite weak, implying that driving the system is necessary to produce significant
temperature anisotropies. However, the absence of forcing, which yields only a small amount of heating, 
makes it easier to precisely identify various waves in the system as was presented in Sec. III. 

\section{Anisotropy of the energy transfer}
The presence of Landau damping can also be seen in the usual wavenumber velocity and magnetic field spectra. 
Considering first simulations with $\beta_0=\beta_{\parallel}=0.8$, Fig. 9 shows the velocity 
spectra with respect to perpendicular and parallel wavenumbers $k_\perp$, $k_\parallel$ which are defined as
$E^U = \int E^u (k_\perp) dk_\perp = \int E^u (k_\parallel) dk_\parallel$. With respect to $k_\perp$, the spectra for Hall-MHD and
FLR-Landau fluid are almost identical (Fig. 9 left) whereas with respect to $k_\parallel$ (Fig. 9 right), the spectra of 
Landau fluid are much steeper. Landau damping therefore significantly inhibits the parallel transfer. Even though 
low resolution does not allow to precisely identify the slopes of the spectra, three straight lines were added to figures 
and correspond to power law solutions $k^{s}$, where $s=-3/2,-5/3$ and $-7/3$. For $E^u(k_\perp)$, 
the closest spectral index value appears to be $-5/3$, the spectral range being however quite limited.
The same conclusion with the inhibition of parallel transfer is also obtained for the magnetic field spectra, which are 
almost identical to the velocity spectra and are shown in Fig. 10.
\begin{figure}
$$\includegraphics[width=0.5\textwidth]{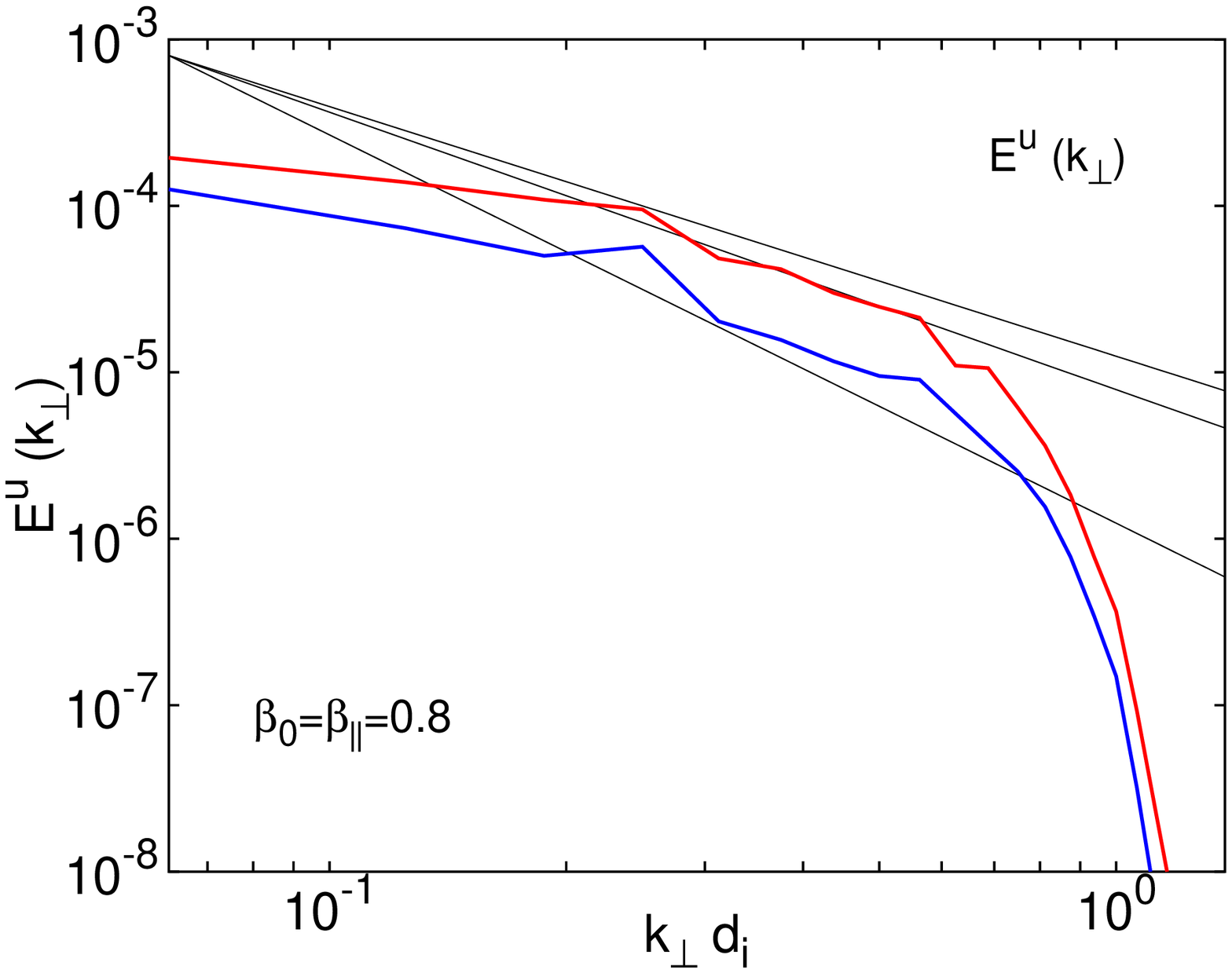}
\includegraphics[width=0.5\textwidth]{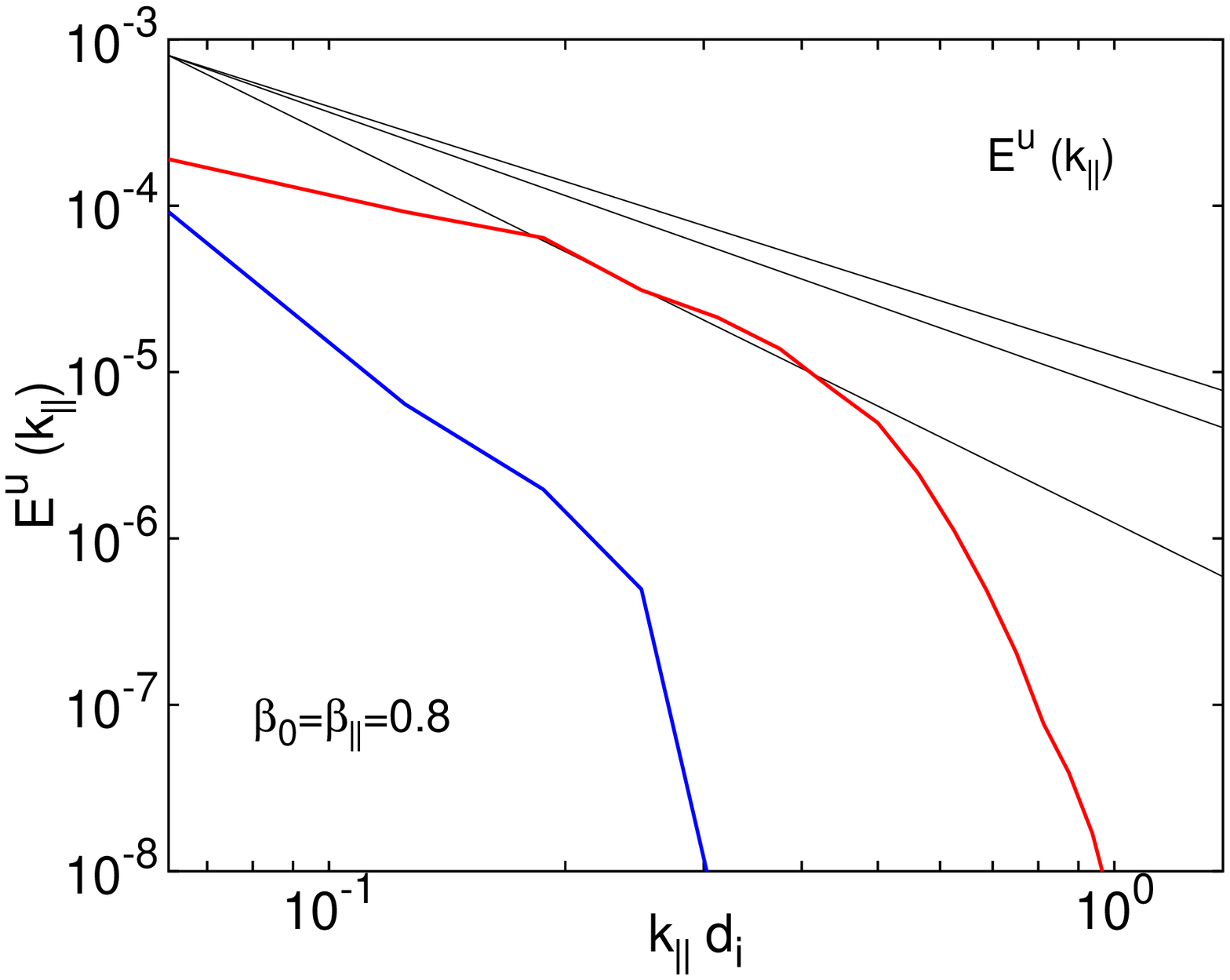}$$
\caption{Velocity spectra for Hall-MHD (red) and Landau fluid (blue) with respect to perpendicular wavenumber $E^u (k_\perp)$ 
(left) and with respect to parallel wavenumber $E^u (k_\parallel)$ (right), for $\beta_0=\beta_\parallel=0.8$. 
Spectra were taken at time $t=5248$. Straight lines correspond to $k^{-3/2}, k^{-5/3}$ and $k^{-7/3}$.
The figure shows that spectra with respect to $k_{\parallel}$ are much steeper in Landau fluid simulation, which is a result
of Landau damping.}
\end{figure}
\begin{figure}
$$\includegraphics[width=0.5\textwidth]{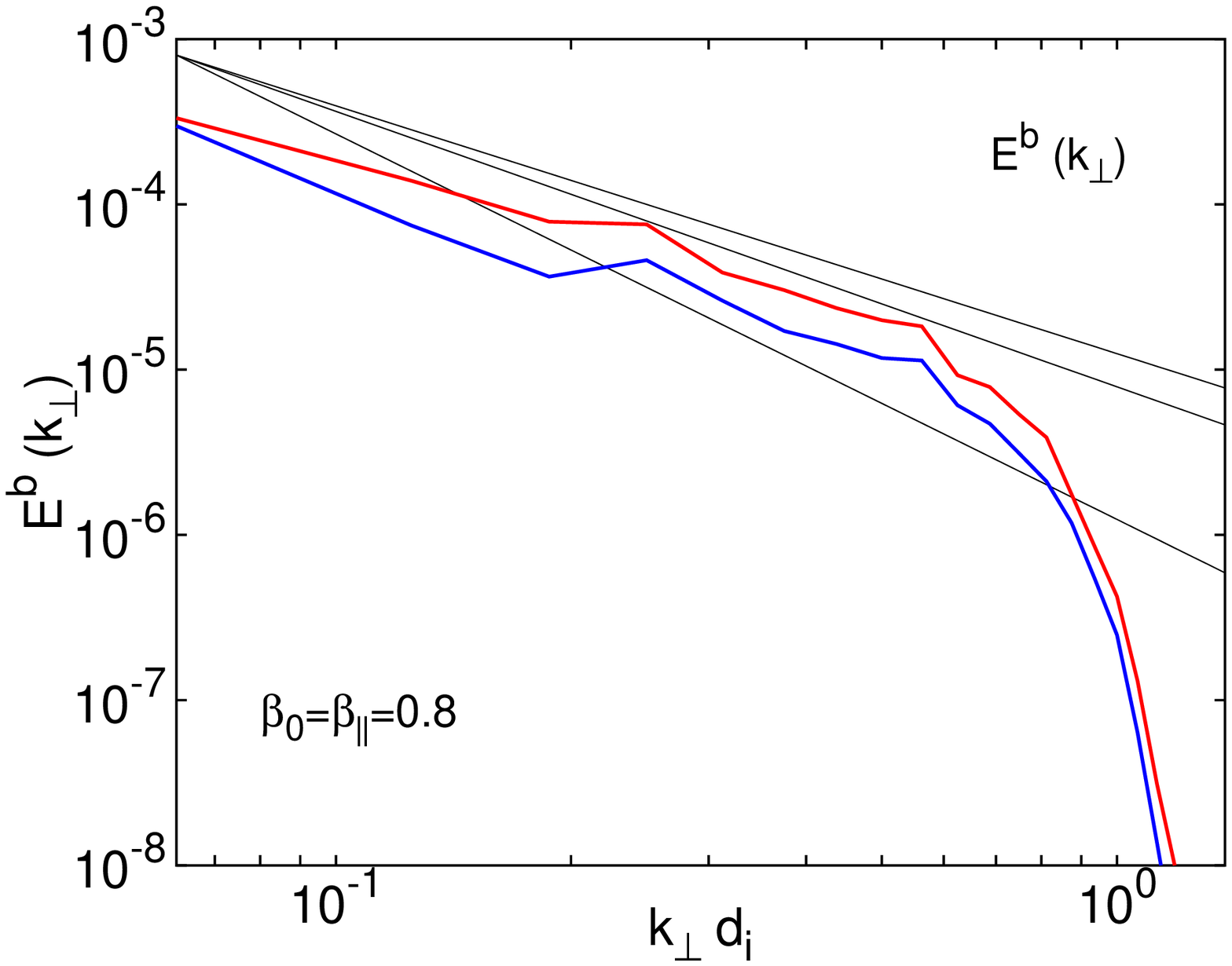}
\includegraphics[width=0.5\textwidth]{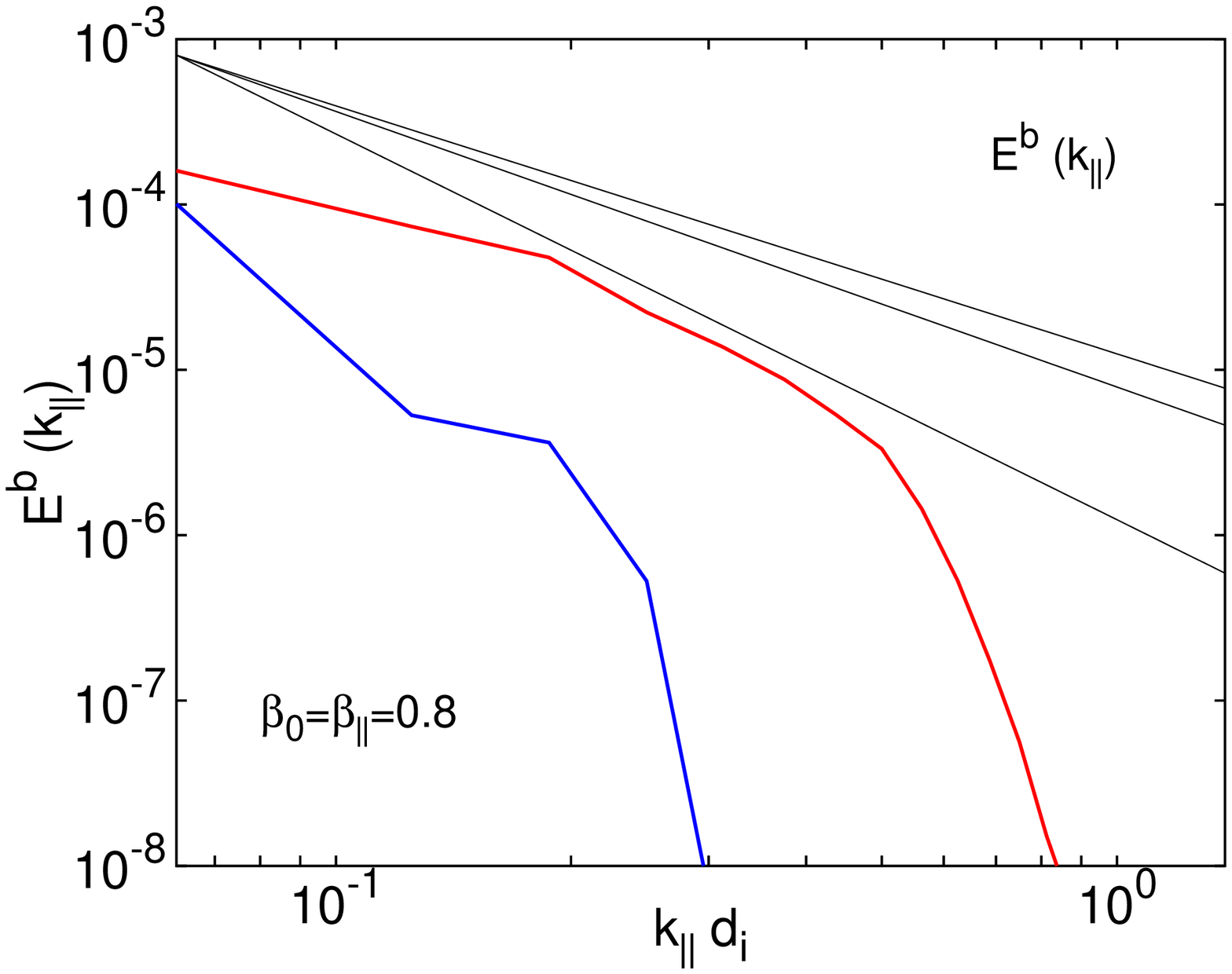}$$
\caption{Magnetic field spectra for Hall-MHD (red) and Landau fluid (blue), for $E^b (k_\perp)$ (left) and $E^b (k_\parallel)$ (right)
when $\beta_0=\beta_\parallel=0.8$. Spectra were taken at time $t=5248$.}
\end{figure}
In contrast, for simulations with $\beta_0=\beta_\parallel=0.1$ the parallel spectrum $E (k_\parallel)$ 
displays a similar behavior for Hall-MHD and Landau fluids.  
The velocity spectra are shown in Fig. 11 and the magnetic field spectra are shown in Fig. 12.
This is consistent with results presented in the previous section where it was shown
that the Landau damping is responsible for significant reduction of compressibility 
for simulations with $\beta_0=\beta_\parallel=0.8$, whereas for simulations with $\beta_0=\beta_\parallel=0.1$, the 
Landau damping was much weaker and the reduction of compressibility almost negligible. 
\begin{figure}
$$\includegraphics[width=0.5\textwidth]{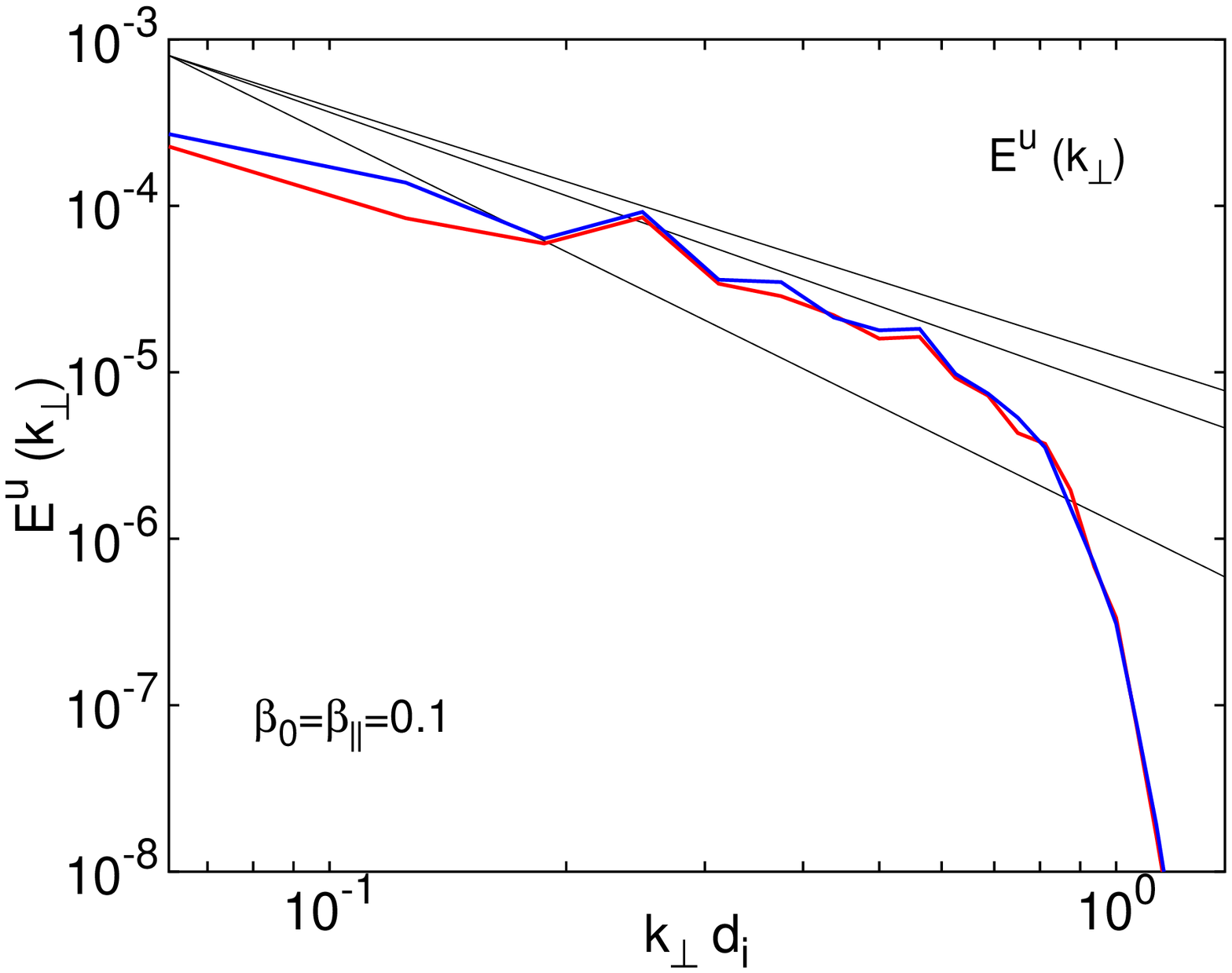}
\includegraphics[width=0.5\textwidth]{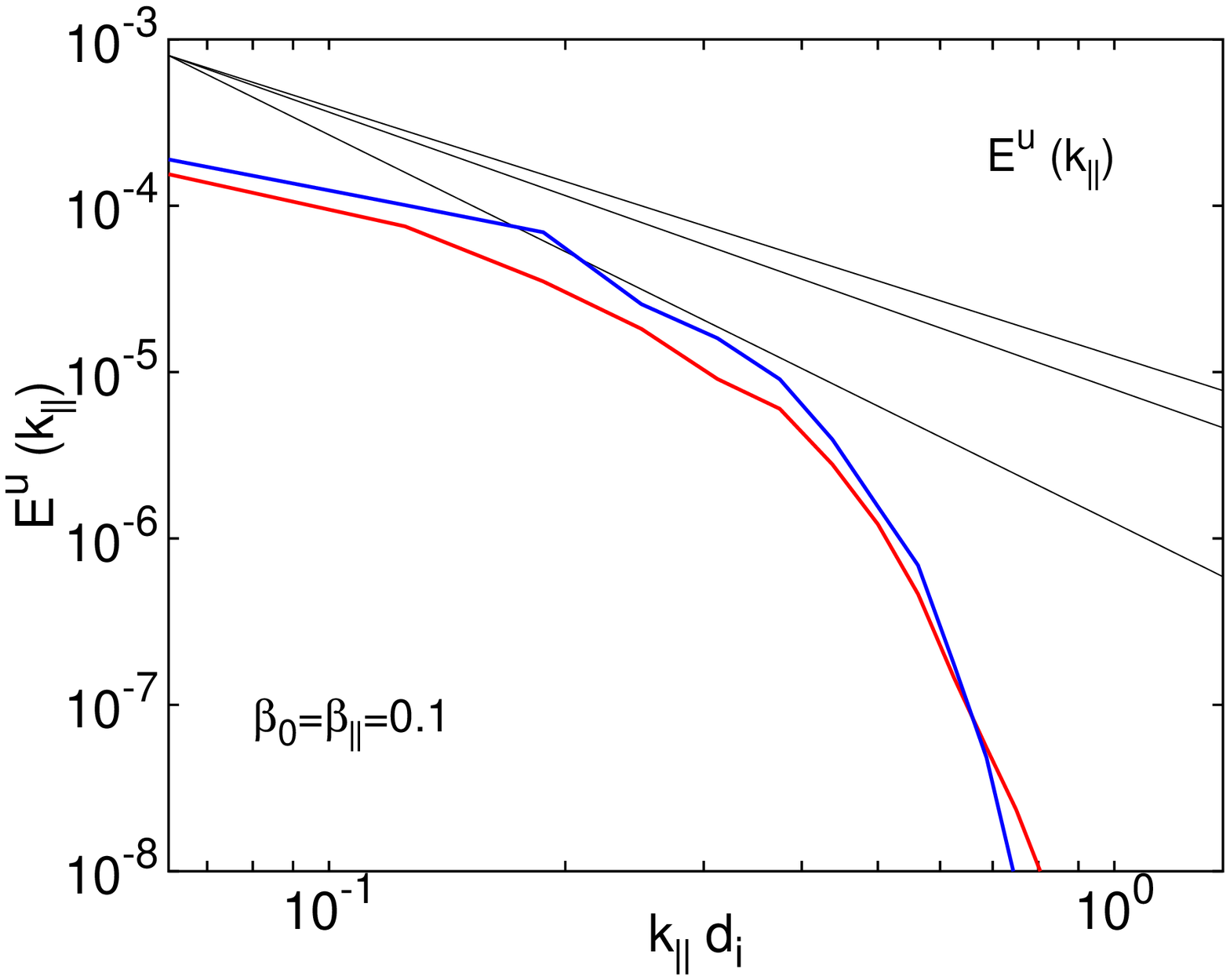}$$
\caption{Velocity spectra for Hall-MHD (red) and Landau fluid (blue) with respect to perpendicular wavenumber $E^u (k_\perp)$ 
(left) and with respect to parallel wavenumber $E^u (k_\parallel)$ (right) for $\beta_0=\beta_\parallel=0.1$. Spectra 
were taken at time $t=5248$. Straight lines correspond to $k^{-3/2}, k^{-5/3}$ and $k^{-7/3}$.}
\end{figure}
\begin{figure}
$$\includegraphics[width=0.5\textwidth]{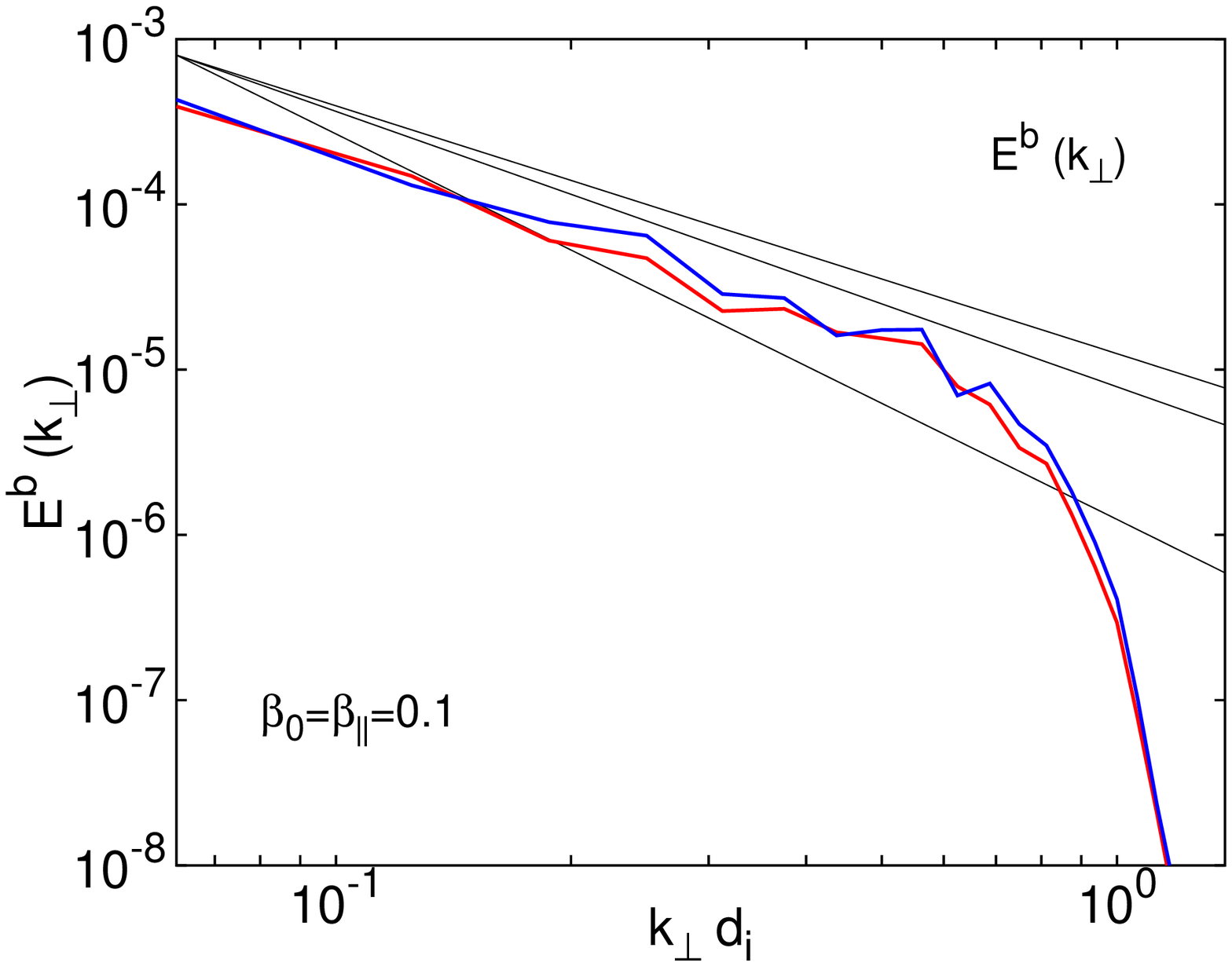}
\includegraphics[width=0.5\textwidth]{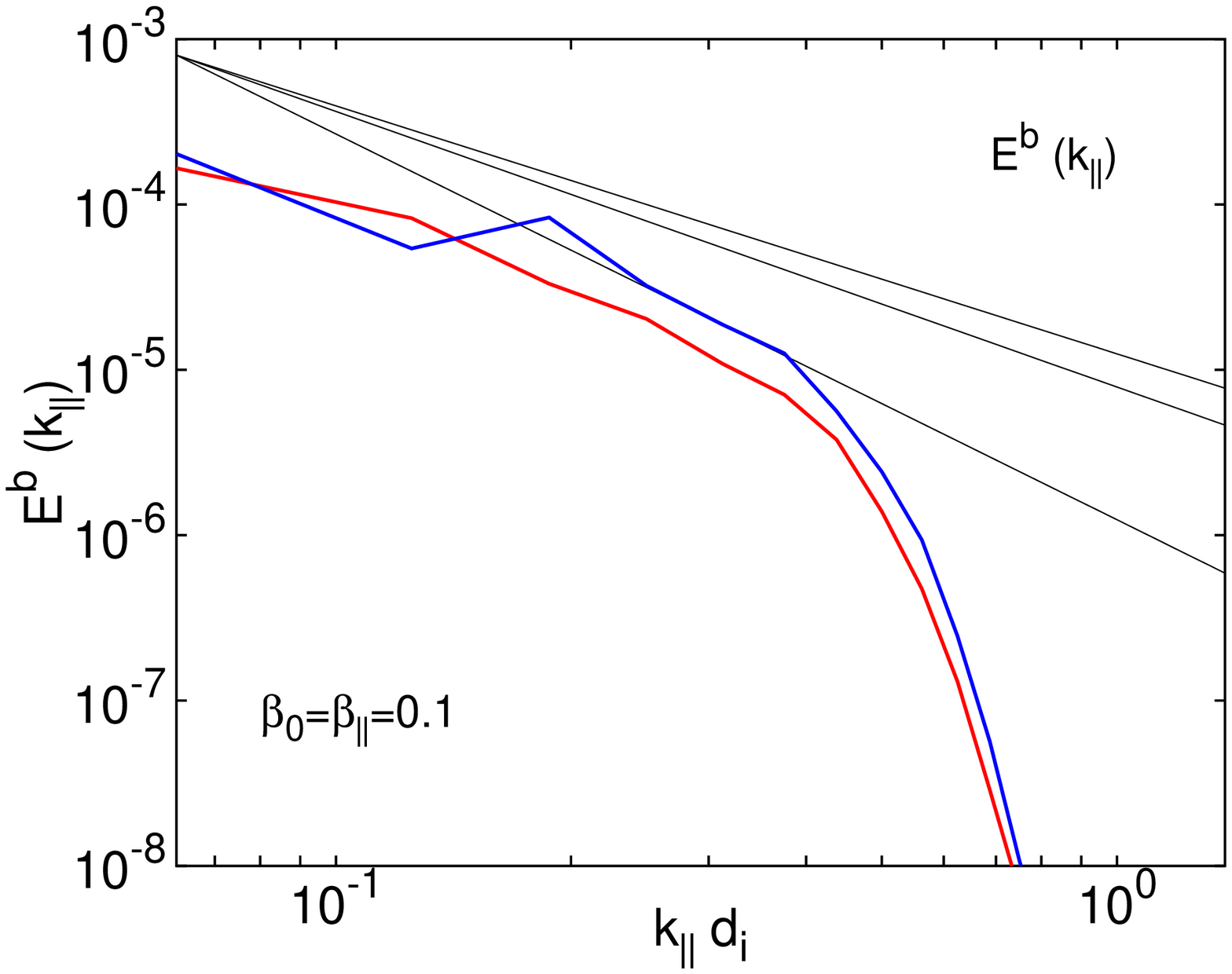}$$
\caption{Magnetic field spectra for Hall-MHD (red) and Landau fluid (blue), for $E^b (k_\perp)$ (left) and $E^b (k_\parallel)$ (right) 
when $\beta_0=\beta_\parallel=0.1$. Spectra were taken at time $t=5248$. Spectra for Hall-MHD and Landau fluid model are again almost identical.}
\end{figure}

It is useful to compare our compressible simulations to incompressible simulations. 
Naturally, our compressible Hall-MHD code cannot be run in an incompressible regime, 
for which the turbulent sonic Mach number $M_s\rightarrow 0$ and the sound speed $c_s\rightarrow \infty$.
Nevertheless, incompressible MHD simulations of decaying turbulence were performed, for example, by Bigot et al. \cite{Bigot2008}. 
These simulations showed that the combined
velocity and magnetic field spectra (they used Els\"asser variable $\boldsymbol{z}^+$) are much steeper with respect to $k_{\parallel}$ 
than with respect to $k_\perp$, if the ambient magnetic field is sufficiently strong. Our compressible simulations presented 
here have initially $\langle\bb^2\rangle^{1/2}/B_0=1/8$ and can therefore be compared to incompressible MHD 
simulations of Bigot et al. \cite{Bigot2008} (their Fig. 17 c,d), where this ratio is $1/5$ and $1/15$, respectively.  
Considering compressible Hall-MHD, our simulations showed that spectra with respect to $k_\parallel$ (e.g. Fig. 9 right, red line)
are steeper than spectra with respect to $k_\perp$ (Fig. 9 left, red line). However, these parallel spectra 
are nowhere near as steep as the parallel spectra of Bigot et al. \cite{Bigot2008}. 
Interestingly, their parallel spectra more resemble our parallel spectra for Landau fluid model (Fig. 9 right, blue line),
where the Landau damping was strong ($\beta_\parallel =0.8$). 
These results show that magnetosonic waves (which are not present in an incompressible MHD description)
play an important role in regulating the parallel energy cascade.   
Steeper parallel spectra in Landau fluid model are therefore a consequence of damping of slow magnetosonic waves by Landau damping.

\section{Conclusion}
We have presented the first three-dimensional fluid simulations of decaying turbulence in a collisionless plasma in conditions close
to the solar wind. For this purpose, we used the FLR-Landau fluid model that extends compressible Hall-MHD by incorporating 
low-frequency kinetic effects such as Landau damping and finite Larmor radius corrections. 
It was shown that in spite of the turbulent regime, it is possible to precisely identify linear waves present in the system.
Comparisons between compressible Hall-MHD and FLR-Landau fluid model showed that when beta is not too small, 
linear Landau damping yields strong damping of slow magnetosonic waves in Landau fluid simulations. These waves are indeed damped in kinetic 
theory described by the Vlasov-Maxwell equations but not in compressible MHD and Hall-MHD descriptions,
which overestimate compressibility and parallel transfer in modeling weakly collisional plasmas.
The FLR-Landau fluid model can therefore be useful for simulating the solar wind, which is typically found to be only weakly compressible.

\section*{Acknowledgements}
The support of INSU-CNRS ``Programme Soleil-Terre'' is acknowledged. Computations were performed on the Mesocentre SIGAMM machine 
hosted by the Observatoire de la C{\^ o}te d'Azur (OCA) and on the JADE cluster of the CINES computational
facilities. PH was supported by an OCA Poincar\'e fellowship. The work of DB was supported by the European Community under the 
contract of Association between EURATOM and ENEA. The views and opinions expressed herein do not necessarily reflect
those of the European Commission.

\section*{APPENDIX}
To clearly understand how the Landau damping acts in the present system (1)-(12), it is useful 
to solve dispersion relations for linear waves propagating in parallel direction to the ambient magnetic field. 
A detailed analysis of linear waves for various propagation angles was elaborated by Passot and Sulem \cite{PassotSulem2004}.
To simplify the analytic expressions, we define the proton temperature anisotropy as $T_{\perp p}^{(0)}/T_{\parallel p}^{(0)}\equiv a_p$, 
and the normalized electron temperature as $T_e^{(0)}/T_{\parallel p}^{(0)}\equiv \tau$. It can be shown that for parallel propagation angle, 
the Landau fluid model contains four dispersive Alfv\'en waves. Two waves obey the dispersion relation
which can be expressed as
\begin{equation}
\omega = \pm \frac{k^2}{2R_i} \left[ 1+\beta_\parallel \left(1-\frac{a_p}{2}\right) \right]
+ k\sqrt{ 1+ \frac{\beta_\parallel}{2}(a_p-1) + \left(\frac{k}{2R_i} \right)^2 
 \left[ 1-\beta_\parallel \left( 1- \frac{a_p}{2}\right) \right]^2 }, \label{eq:AlfvenGen}
\end{equation}
with another two solutions obtained by substituting $\omega$ with $-\omega$. Obviously,  
these Alfv\'en waves are independent of the electron temperature $\tau$, which is a consequence 
of electrons being modeled as isothermal. For a more general Landau fluid model which contains 
evolution equation for electron pressures and heat fluxes, the electron temperature $\tau$ enters 
the dispersion relation for Alfv\'en waves. The solutions (\ref{eq:AlfvenGen}) can become 
imaginary, if the expression under the square root becomes negative. At large scales (when $1/R_i\rightarrow 0$)
the condition $1+\beta_\parallel(a_p-1)/2<0$ represents the well known criterion for fire-hose instability 
(see, for example, Ferri\`ere and Andr\'e \cite{FerriereAndre2002}). The Hall term and FLR corrections modify
the instability criterion.  
For isotropic temperatures ($a_p=1$), the solution (\ref{eq:AlfvenGen}) naturally 
collapses to the solution (\ref{eq:zeroAlfven}). 
The four Alfv\'en waves (\ref{eq:AlfvenGen}) can be 
eliminated from the general dispersion relation and this yields a complex polynomial of 6th order 
in frequency $\omega$. Solutions of this polynomial represent 3 forward and 3 backward propagating waves
which have a negative imaginary part and are therefore damped. Importantly, it is possible to eliminate the  
dependence on $\beta_\parallel$ and wavenumber $k$ and, after applying a substitution  
$\Omega=\omega/(k\sqrt{\beta_\parallel})$, the polynomial of 6th order can be simplified to
\begin{eqnarray}
&&\Omega^6 + \Omega^5 i \sqrt{\pi}\frac{-4+3\pi}{-16+6\pi}
+\Omega^4 \frac{-\pi(14+3\tau)+24+8\tau}{-16+6\pi} + 
\Omega^3 i \sqrt{\pi}\frac{\tau-\pi(9+3\tau)/4}{-8+3\pi} \nonumber \\
&&+\Omega^2 \frac{\pi(15+5\tau)/4-2}{-8+3\pi} +
\Omega^1 i \sqrt{\pi}\frac{(5+3\tau)/2}{-8+3\pi} 
- \frac{1+\tau}{-8+3\pi} =0. \label{eq:Omega}
\end{eqnarray} 
Because this polynomial in $\Omega$ does not depend on $\beta_\parallel$ or $k$, the substitution
implies that all 6 waves are linear with $k$ and $\sqrt{\beta_\parallel}$. The polynomial (\ref{eq:Omega}) 
has to be solved numerically for a given value of $\tau$. 
The simulations presented here use $\tau=1$ and numerically solving polynomial (\ref{eq:Omega}) 
yields $\Omega= \pm 1.48106 -i 0.36117$, $\Omega= \pm 0.65467 -i 0.88285$ and 
$\Omega=\pm 0.55104 -i 0.44311 $. These six waves therefore satisfy
\begin{eqnarray}
&& \omega = k\sqrt{\beta_\parallel} ( \pm 1.48106 -i 0.36117 ) \label{eq:sound1},\\
&& \omega = k\sqrt{\beta_\parallel} ( \pm 0.65467 -i 0.88285 ) \label{eq:sound2},\\
&& \omega = k\sqrt{\beta_\parallel} ( \pm 0.55104 -i 0.44311 ) \label{eq:sound3}.
\end{eqnarray}
The least damped solution (\ref{eq:sound1}) represents the sound wave. 
The solutions of Landau fluid model (1)-(12) for parallel propagation angle are therefore 4 Alfv\'en waves 
(\ref{eq:AlfvenGen}), 2 sound waves (\ref{eq:sound1}) and 4 waves (\ref{eq:sound2}), (\ref{eq:sound3}), which are highly damped.
These 4 waves (\ref{eq:sound2}), (\ref{eq:sound3}) do not have an analogy in Hall-MHD description and correspond
to solutions of kinetic Maxwell-Vlasov description, which contains an infinite number of highly damped solutions. 
Interestingly, the last solution (\ref{eq:sound3}) is not dependent on the value of $\tau$ and it can be expressed analytically as  
$\Omega=\pm \sqrt{8-\pi}/4 -i\sqrt{\pi}/4$. After eliminating these waves from eq. (\ref{eq:Omega}), the polynomial which 
contains the sound waves (\ref{eq:sound1}) and solutions (\ref{eq:sound2}) is now of 4th order in $\Omega$ and expressed as
\begin{eqnarray}
\Omega^4 +i\frac{2\sqrt{\pi}}{-8+3\pi}\Omega^3 -\frac{1}{2}\frac{9\pi-16 +(-8+3\pi)\tau}{-8+3\pi}\Omega^2
-i\frac{\sqrt{\pi}(3+\tau)}{-8+3\pi}\Omega +\frac{2(1+\tau)}{-8+3\pi}=0. \label{eq:Omegafin}
\end{eqnarray}
It is of course possible to use Ferrari-Cardano's relations to solve this polynomial analytically, 
the final result is however too complicated and it is still more convenient to solve (\ref{eq:Omegafin}) numerically for a given 
value of $\tau$. 

If this wave analysis is repeated with the model (1)-(10) with heat flux equations 
$q_\parallel=0$, $q_\perp=0$, the same dispersion relation (\ref{eq:AlfvenGen}) for Alfv\'en waves is obtained. 
However, the only other solutions present in the parallel direction are
\begin{equation}
\omega = \pm k\sqrt{\frac{\beta_\parallel}{2}(3+\tau)}.
\end{equation} 
These waves have frequencies which are purely real and correspond to undamped sound waves of double adiabatic 
model with isothermal electrons. 

For completeness, considering perpendicular propagation, it can be shown that the heat fluxes
vanish and the solutions of the Landau fluid model with isothermal electrons are undamped magnetosonic waves with the 
dispersion relation expressed as
\begin{equation}
\omega=\pm k \sqrt{1+\beta_\parallel \left( a_p+\frac{\tau}{2}\right) + \left( \frac{a_p\beta_\parallel k}{4 R_i} \right)^2 }.
\end{equation}



\hyphenation{Post-Script Sprin-ger}


\end{document}